\documentclass[aps,
english,preprintnumbers,nofootinbib,
twocolumn,physics]{revtex4-1}

\usepackage{amsfonts,amsmath,amssymb}
\usepackage{graphicx}
\usepackage[utf8]{inputenc}
\usepackage{hyperref}
\usepackage{babel}

\usepackage{xcolor}

\newcommand\be{\begin{equation}}
\newcommand\ee{\end{equation}}
\newcommand\bea{\begin{eqnarray}}
\newcommand\eea{\end{eqnarray}}

\begin{document}

\title{Krylov Distribution}
\author{Mohsen Alishahiha and  Mohammad Javad Vasli}
\email{alishah@ipm.ir, vasli@ipm.ir}
\affiliation{School of Quantum Physics and Matter, Institute for Research in Fundamental Sciences (IPM),\\
	P.O. Box 19395-5531, Tehran, Iran\\} 


\begin{abstract}
We introduce the Krylov distribution $\mathcal{D}(\xi)$, a static Krylov-space diagnostic that characterizes how inverse-energy response is organized in Hilbert space. The central object is the resolvent-dressed state $(H-\xi)^{-1}|\psi_0\rangle$, whose decomposition in the Krylov basis generated from a reference state defines a normalized distribution over Krylov levels. Unlike conventional spectral functions, which resolve response solely along the energy axis, the Krylov distribution captures how the resolvent explores the dynamically accessible subspace as the spectral parameter $\xi$ is varied. Using asymptotic analysis, exact results in solvable models, and numerical studies of an interacting spin chain, we identify three universal regimes: saturation outside the spectral support, extensive growth within continuous spectra, and sublinear or logarithmic scaling near spectral edges and quantum critical points. We further show that fidelity susceptibility and the quantum geometric tensor admit natural decompositions in terms of Krylov-resolved resolvent amplitudes.

\end{abstract}

\maketitle

\section{Introduction}

How quantum states explore Hilbert space under the action of a Hamiltonian is a central question in quantum many-body physics. Phenomena such as thermalization~\cite{Srednicki:1994mfb,Deutsch:1991}, information scrambling~\cite{Shenker2014}, quantum chaos~\cite{Bohigas1984}, and quantum criticality~\cite{Polkovnikov2011} are all, in different ways, governed by how an initially simple state becomes distributed over increasingly complex structures in Hilbert space. Developing organizing principles for this exploration remains a key theoretical challenge.

Krylov-space methods provide a particularly powerful framework for addressing this problem. Starting from a reference state $|\psi_0\rangle$, the Krylov construction generates an ordered orthonormal basis by repeated action of the Hamiltonian, in which $H$ assumes a tridiagonal (Jacobi) form~\cite{Lanczos1950,Haydock1980,ParkLight1986,viswanath1994recursion,viswanath1994}. This representation endows Hilbert space with an emergent one-dimensional geometry encoded in the Lanczos coefficients.

Most existing applications of Krylov-space methods focus on unitary time evolution. In this setting, the spreading of a time-evolved state $|\psi(t)\rangle = e^{-iHt}|\psi_0\rangle$ along the Krylov chain is quantified by the Krylov complexity ~\cite{Parker:2018yvk,Rabinovici:2020ryf,Balasubramanian:2022tpr,PhysRevResearch.4.013041,Barbon:2019wsy,Alishahiha:2022anw}. Krylov complexity has proven to be a sensitive diagnostic of operator growth, quantum chaos, and universality across a wide range of quantum systems, including spin chains, random matrix models, conformal field theories, and holographic setups
\cite{Dymarsky:2021bjq,Bhattacharjee:2022vlt,Avdoshkin:2022xuw,Camargo:2022rnt,Vasli:2023syq,Imani:2025etp,Rabinovici:2022beu,Scialchi:2023bmw,Trigueros:2021rwj,Espanol:2022cqr,Erdmenger:2023wjg,Huh:2023jxt,Camargo:2024deu,Nandy:2024wwv,Bhattacharjee:2024yxj,Balasubramanian:2024ghv,Baggioli:2024wbz,Alishahiha:2024vbf,FarajiAstaneh:2025rlc,Bhattacharya:2024szw,Huh:2024ytz,Baggioli:2025ohh}, as reviewed in Refs.~\cite{Nandy:2024evd,Rabinovici:2025otw}. These developments have clarified how dynamical complexity is governed by the growth of Lanczos coefficients and the emergent geometry of Krylov space.

Many physically significant phenomena, however, are not primarily dynamical but instead arise from static or quasi-static responses to external parameters. Examples include adiabatic deformations, geometric phases, inverse-gap physics, and the breakdown of adiabaticity near quantum critical pointss~\cite{ProvostVallee1980,Zanardi2007,GuShiJian:2007dsd,Kolodrubetz2017,Damski:2013rup}. Such phenomena are naturally encoded not in the time-evolution operator $e^{-iHt}$, but in the resolvent $(H-\xi)^{-1}$, which weights virtual excitations according to their inverse energy separation. This raises a natural question: how does inverse-energy response explore Krylov space, and what does this reveal about spectral structure?

In this work, we introduce a Krylov-space framework tailored specifically to resolvent physics. Our central object is the resolvent-dressed state
\begin{equation}
    |\psi(\xi)\rangle = (H-\xi)^{-1}|\psi_0\rangle,
\end{equation}
and its decomposition in the Krylov basis generated from the reference state $|\psi_0\rangle$. From this decomposition, we define the Krylov distribution $D(\xi)$, a static, resolvent-based diagnostic that characterizes how inverse-energy response is spatially distributed along the Krylov chain. Rather than extracting spectral functions, which resolve response solely along the energy axis, the Krylov distribution organizes resolvent response in terms of its structure within the dynamically accessible subspace associated with $|\psi_0\rangle$.

Concretely, writing $|\psi(\xi)\rangle=\sum_n \psi_n(\xi)|n\rangle$ with $\psi_n(\xi)=\langle n|(H-\xi)^{-1}|\psi_0\rangle$, we define a normalized probability distribution
\begin{equation}
P_n(\xi)=\frac{|\psi_n(\xi)|^2}{\sum_\ell|\psi_\ell(\xi)|^2},
\end{equation}
and its first moment,
\begin{equation}
\mathcal{D}(\xi)=\sum_n n\,P_n(\xi),
\end{equation}
which we refer to as the Krylov distribution. While the amplitudes $\psi_n(\xi)$ appear naturally in the recursion method and Green’s-function theory~\cite{Haydock1980,viswanath1994recursion}, the normalized quantity $\mathcal{D}(\xi)$, its scaling behavior, and its interpretation as a measure of inverse-energy spreading in Krylov space are new.

Our key conceptual advance is to elevate the resolvent-dressed state from an auxiliary object used to compute spectral functions to a primary diagnostic whose spatial organization in Krylov space encodes spectral regime, scaling behavior, and static response properties.

The Krylov distribution plays a role directly analogous to Krylov complexity, with the spectral parameter $\xi$ replacing time as the control variable. Whereas Krylov complexity tracks how quantum states spread dynamically, $\mathcal{D}(\xi)$ characterizes how inverse-energy weight is distributed across Krylov space as $\xi$ probes different spectral regions. In this way, it provides access to geometric information about spectral organization that is invisible to purely dynamical diagnostics.

Using asymptotic analysis, exact solutions, and numerical simulations, we show that $\mathcal{D}(\xi)$ exhibits distinct and universal behavior depending on the location of $\xi$ relative to the spectrum. When $\xi$ lies outside the spectral support or is separated from it by a finite gap, resolvent amplitudes are exponentially localized in Krylov space, leading to saturation of $\mathcal{D}(\xi)$ in the thermodynamic limit. When $\xi$ lies within a continuous part of the spectrum associated with an absolutely continuous spectral measure, the amplitudes remain extended in an averaged sense, resulting in extensive growth of $\mathcal{D}(\xi)$. Near spectral edges and quantum critical points, sublinear or logarithmic scaling emerges, reflecting singular features of the density of states and the asymptotic structure of the Lanczos coefficients.

These behaviors are established through a combination of orthogonal polynomial techniques, spectral analysis of Jacobi operators, exact results in analytically tractable models, and numerical studies of interacting systems. In particular, we obtain closed-form results for three paradigmatic Krylov chains: models with constant Lanczos coefficients and bounded continuous spectra; the displaced harmonic oscillator with square-root-growing coefficients and an unbounded discrete spectrum; and the $SU(1,1)$ chain with linearly growing coefficients, which captures the asymptotic Krylov structure of systems with continuous spectra, including maximally chaotic models.

Finally, we show that the Krylov distribution provides a natural bridge between spectral structure and static response. Fidelity susceptibility, the quantum geometric tensor, and higher inverse-gap moments admit transparent decompositions in terms of Krylov-resolved resolvent amplitudes, extending geometric response theory beyond energy eigenstates to arbitrary reference states.

This work is organized as follows. In Section~\ref{sec:K-D}, we introduce the Krylov resolvent amplitudes and define the Krylov distribution, deriving its basic properties and connections to fidelity susceptibility and quantum geometry. Section~\ref{sec:Large-N} establishes the asymptotic behavior of $\mathcal{D}(\xi)$ in the thermodynamic limit, distinguishing between gapped, continuous-spectrum, and critical regimes. Section~\ref{sec:Three-Model} presents exact solutions in three solvable models, while Section~\ref{sec:Ising} contains numerical studies of the mixed-field Ising model. We conclude in Section~\ref{sec:conclusion}, with technical details provided in the Appendices.

\section{Krylov Resolvent Amplitudes and the Krylov Distribution}
\label{sec:K-D}

This section constitutes the conceptual core of the paper. Here we introduce a static,
resolvent-based diagnostic formulated in Krylov space, which might be used to probe spectral structure,
quantum geometry, and criticality in a unified manner. Our central objects are the
Krylov resolvent amplitudes and the associated Krylov distribution.
They provide a Krylov-space decomposition of resolvent-dressed states and serve as
static analogues of Krylov complexity, encoding how different spectral regions are
accessed by a given reference state.

We begin by briefly recalling the construction of the Krylov basis via the Lanczos
algorithm. Consider a quantum system with a time-independent Hamiltonian $H$ acting
on a Hilbert space $\mathcal{H}$ of dimension $\mathcal{D}$. Given a normalized reference
state $|\psi_0\rangle$, the Krylov space generated by $H$ and $|\psi_0\rangle$ is
\begin{equation}\label{Krylov-space}
\mathcal{K} = \mathrm{span}\{|\psi_0\rangle, H|\psi_0\rangle, H^2|\psi_0\rangle, \dots\},
\end{equation}
with dimension $d_\psi \le \mathcal{D}$.

The Lanczos algorithm \cite{Lanczos1950,viswanath1994recursion} constructs an orthonormal
basis $\{|n\rangle\}_{n=0}^{d_\psi-1}$ for $\mathcal{K}$ by iteratively orthogonalizing
$H^n|\psi_0\rangle$. Starting from $|0\rangle=|\psi_0\rangle$ and defining
$|{-1}\rangle=0$, the recursion takes the form
\begin{align}\label{GS}
|\widehat{n+1}\rangle &= (H-a_n)|n\rangle - b_n|n-1\rangle,\\
|n+1\rangle &= \frac{|\widehat{n+1}\rangle}{b_{n+1}},
\end{align}
with Lanczos coefficients
\begin{equation}
a_n=\langle n|H|n\rangle, \qquad
b_{n+1}=\sqrt{\langle\widehat{n+1}|\widehat{n+1}\rangle}.
\end{equation}
In this basis, the Hamiltonian becomes a symmetric tridiagonal Jacobi matrix,
\begin{equation}
H=\sum_{n=0}^{d_\psi-1} a_n |n\rangle\langle n|
+ \sum_{n=0}^{d_\psi-2} b_{n+1} (|n\rangle\langle n+1|+|n+1\rangle\langle n|),
\end{equation}
making explicit the interpretation of Krylov space as a one-dimensional chain with
site-dependent hopping amplitudes.

Much of the recent literature has focused on unitary time evolution
$|\psi(t)\rangle=e^{-iHt}|\psi_0\rangle$, whose Krylov amplitudes
$\phi_n(t)=\langle n|\psi(t)\rangle$ obey a tight-binding Schrödinger equation along
the Krylov chain
\begin{equation}\label{eq:dotphi}
i\dot\phi_n(t)=a_n\phi_n(t)+b_{n+1}\phi_{n+1}(t)+b_n\phi_{n-1}(t)\,.
\end{equation}
The average Krylov position,
\begin{equation}
\mathcal{C}(t)=\sum_{n=0}^{d_\psi-1} n|\phi_n(t)|^2,
\end{equation}
defines the Krylov complexity \cite{Parker:2018yvk}, a dynamical measure of operator or
state growth.

In this work we adopt a complementary,  static  perspective by replacing time
evolution with the resolvent operator
\begin{equation}
R(\xi)=(H-\xi)^{-1},
\end{equation}
where $\xi$ is a real spectral parameter.
Acting on the reference state produces the resolvent-dressed state\footnote{This framework can also be extended to resolvent-dressed operators; see \cite{Lunt:2025dcc} for a related study.}
\begin{equation}
|\psi(\xi)\rangle = R(\xi)|\psi_0\rangle,
\end{equation}
which selectively probes eigenstates with energies near $\xi$.
Expanding this state in the Krylov basis,
\begin{equation}
|\psi(\xi)\rangle = \sum_{n=0}^{d_\psi-1} \psi_n(\xi)|n\rangle,
\end{equation}
defines the Krylov resolvent amplitudes
\begin{equation}
\psi_n(\xi)=\langle n|(H-\xi)^{-1}|\psi_0\rangle.
\end{equation}

These amplitudes satisfy an exact algebraic recursion relation obtained by projecting
$(H-\xi)|\psi(\xi)\rangle=|\psi_0\rangle$ onto the Krylov basis:
\begin{equation}
b_{n+1}\psi_{n+1}(\xi)+(a_n-\xi)\psi_n(\xi)+b_n\psi_{n-1}(\xi)=\delta_{n0},
\label{eq:psi_recursion}
\end{equation}
with $\psi_{-1}=0$. This equation is the static analogue of the Krylov Schrödinger
equation \eqref{eq:dotphi} and is related to it by the Laplace transform~\cite{viswanath1994recursion} (see also
 \cite{Parker:2018yvk, Balasubramanian:2025xkj}),
\begin{equation}\label{Laplace}
\psi_n(\xi+i\eta)
=-i\int_0^\infty dt\,e^{i(\xi+i\eta)t}\phi_n(t),\qquad \eta>0.
\end{equation}

Because the resolvent is not unitary, the amplitudes $\psi_n(\xi)$ do not define a
normalized probability distribution. We therefore introduce the 
Krylov resolvent  probability   distribution
\begin{equation}
P_n(\xi)=\frac{|\psi_n(\xi)|^2}{\mathcal{N}(\xi)},\qquad
\mathcal{N}(\xi)=\sum_{m=0}^{d_\psi-1}|\psi_m(\xi)|^2,
\end{equation}
and define the central diagnostic of this work, the Krylov distribution,
\begin{equation}
\mathcal{D}(\xi)=\sum_{n=0}^{d_\psi-1} n P_n(\xi)
=\frac{\sum_n n|\psi_n(\xi)|^2}{\sum_n|\psi_n(\xi)|^2}.
\end{equation}
This quantity measures the average depth along the Krylov chain reached by
the resolvent-dressed state. It plays a role directly analogous to Krylov complexity,
but with the spectral parameter $\xi$ replacing time as the control variable.

This quantity is bounded, $0\le\mathcal{D}(\xi)\le d_\psi-1$, depends on both the
Hamiltonian and the reference state\footnote{Of course as shown in Appendix~\ref{app:bounds}, the extremal values are generally inaccessible due to constraints from the recursion relation. The actual range depends on system parameters and the spectral position $\xi$.
}, and varies smoothly with $\xi$ away from the
spectrum. Singular behavior in $\mathcal{D}(\xi)$ directly reflects spectral
singularities, making it a sensitive probe of gaps, continua, and critical points.

It is worth noting that the construction we have developed in this section relies on the Krylov subspace $\mathcal{K}$ generated from $|\psi_0\rangle$ via 
unitary evolution \eqref{Krylov-space}. This subspace represents the part of Hilbert space dynamically accessible starting from $|\psi_0\rangle$. An important mathematical fact ensures the consistency of our approach: the resolvent-dressed state $|\psi(\xi)\rangle = (H-\xi)^{-1}|\psi_0\rangle$ necessarily belongs to $\mathcal{K}$. This follows from the Laplace transform relation \eqref{Laplace}. Since $|\psi(t)\rangle = e^{-iHt}|\psi_0\rangle$ remains within $\mathcal{K}$ for all $t$, its Laplace transform $|\psi(\xi)\rangle$ must also lie in $\mathcal{K}$.

This mathematical guarantee means that expanding $|\psi(\xi)\rangle$ in the Krylov basis $\{|n\rangle\}$ from $|\psi_0\rangle$ provides a complete representation of the resolvent-dressed state within the dynamically accessible subspace. The decomposition $\psi_n(\xi) = \langle n|(H-\xi)^{-1}|\psi_0\rangle$ is therefore well-defined and captures the full spectral response as filtered through the Krylov organization from the initial state.

Higher inverse-gap moments arise naturally by considering powers of the resolvent,
$|\psi^{(p)}(\xi)\rangle=R(\xi)^p|\psi_0\rangle$. Using
$\frac{d}{d\xi}R(\xi)=R(\xi)^2$, one finds the exact hierarchy
\begin{equation}
\psi_n^{(p+1)}(\xi)=\frac{1}{p}\frac{d}{d\xi}\psi_n^{(p)}(\xi),
\end{equation}
showing that all higher-order static response functions are generated by derivatives
of the fundamental amplitudes.

This framework connects directly to quantum information geometry. The standard quantum geometric tensor quantifies the sensitivity of the ground state $|E_0\rangle$ to parameter changes through the fidelity susceptibility
\begin{equation}
\chi_F = \langle E_0|(\partial_\lambda H) R_0^2 (\partial_\lambda H)|E_0\rangle,
\end{equation}
where $R_0$ is the reduced resolvent at $E_0$.

From our Krylov-space perspective, this construction extends naturally to arbitrary initial states $|\psi_0\rangle$ that need not be energy eigenstates. For such states, we define a generalized fidelity susceptibility
\begin{equation}
\chi_F^{\psi_0} = \langle \psi_0|(\partial_\lambda H) R_{\psi_0}^2 (\partial_\lambda H)|\psi_0\rangle,
\end{equation}
where $R_{\psi_0}$ is the reduced resolvent excluding $|\psi_0\rangle$. Expanding the resolvent-dressed state $|\psi_{\partial H}\rangle = R_{\psi_0}(\partial_\lambda H)|\psi_0\rangle$ in the Krylov basis generated from $|\psi_0\rangle$ yields a Krylov-resolved decomposition
\begin{equation}
\chi_F^{\psi_0} = \sum_{n \ge 0} |\psi_n^{\partial H}|^2, \qquad
\psi_n^{\partial H} = \langle n | \psi_{\partial H} \rangle.
\end{equation}
This reveals how the parametric sensitivity of $|\psi_0\rangle$ distributes across Krylov layers—the static analogue of the Krylov distribution $\mathcal{D}(\xi)$ introduced earlier. For non-eigenstates, this decomposition is particularly clean as the Krylov subspace is typically non-trivial.

More generally, for multi-parameter Hamiltonians $H(\vec{\lambda})$, we obtain a $\xi$-dependent Krylov quantum geometric tensor
\begin{equation}
Q_{\mu\nu}^{\psi_0}(\vec{\lambda}, \xi) = \sum_{n \ge 0} [\psi_n^{(\mu)}(\vec{\lambda}, \xi)]^* \psi_n^{(\nu)}(\vec{\lambda}, \xi),
\end{equation}
with amplitudes $\psi_n^{(\mu)}(\xi) = \langle n | (H-\xi)^{-1} \partial_\mu H | \psi_0 \rangle$. This provides both spectral resolution through $\xi$ and distribution across Krylov layers through $n$, offering a unified perspective on quantum geometry. The complete formulation, including subtleties for eigenstates is developed in Appendix~\ref{app:fidelity}.

Finally, it is worth noting that since our analysis is formulated in terms of the resolvent rather than unitary time evolution, one could consider an alternative Krylov construction adapted directly to inverse powers of the Hamiltonian. In addition to the standard Krylov basis generated by repeated applications of $H$, one may define a resolvent Krylov basis generated by powers of $H^{-1}$ acting on the reference state. This basis spans the same spectral subspace but effectively reweights energy scales according to $E \mapsto E^{-1}$, thereby emphasizing low-energy structure while suppressing high-energy contributions. Krylov distributions defined in this basis probe complementary aspects of the spectrum and provide an alternative, resolvent-adapted characterization of inverse-gap physics. Although in this paper we work primarily with the standard Krylov space, for completeness we describe the construction of the resolvent Krylov basis and the associated recursion relations in Appendix~\ref{app:res_krylov}.


\section{Krylov Resolvent Amplitudes and the Thermodynamic Limit}
\label{sec:Large-N}

In the previous section we introduced the Krylov distribution as a static
diagnostic characterizing how spectral weight is distributed along the Krylov
chain. Its evaluation relies on the Krylov resolvent amplitudes
$\psi_n(\xi)$, which therefore play a central role in the analysis.

In this section we study the general properties of the Krylov resolvent
amplitudes with the goal of identifying universal features of the Krylov
distribution in the thermodynamic limit. Here the thermodynamic limit refers to
systems with a large effective Krylov dimension, characterized by a large number
$N$ of Lanczos iterations retained in the Krylov construction. Physically, $N$
should be understood as a Krylov-space cutoff rather than a microscopic system
size, although in many-body systems the two are closely related.

A key observation is that, for large $N$, the discrete recursion relations
satisfied by the Krylov resolvent amplitudes admit a controlled continuum and
asymptotic description. The amplitudes $\psi_n(\xi)$ are naturally connected to
the resolvent of Jacobi operators and to the associated theory of orthogonal
polynomials \cite{Chihara1978,Szegoe,Deift,SimonOPRL}. This connection allows one
to exploit powerful results from spectral theory and orthogonal polynomial
asymptotics to characterize the large-$n$ behavior of $\psi_n(\xi)$ and, in turn,
the scaling of the Krylov distribution. Related applications of orthogonal
polynomials in the context of Krylov complexity have appeared in
Refs.~\cite{Muck:2022,Kar:2022,Muck:2024,Chhetriya:2025,
Alishahiha:2024vbf,
Balasubramanian:2025xkj,Qu:2025lgo}.

Inserting a complete set of energy eigenstates,
$H|E_\alpha\rangle = E_\alpha |E_\alpha\rangle$, the Krylov resolvent amplitudes
admit the spectral decomposition
\begin{equation}
\psi_n(\xi)
= \sum_\alpha
\frac{\langle n | E_\alpha\rangle \langle E_\alpha | \psi_0\rangle}
     {E_\alpha - \xi}.
\end{equation}
Introducing the normalized overlaps
\begin{equation}
Q_n(E_\alpha)
\equiv
\frac{\langle n | E_\alpha\rangle}
     {\langle \psi_0 | E_\alpha\rangle},
\end{equation}
we may write
\begin{equation}
\psi_n(\xi)
= \sum_\alpha
\frac{Q_n(E_\alpha)\, d\mu_0(E_\alpha)}{E_\alpha - \xi},
\end{equation}
where
\begin{equation}
d\mu_0(E)
=
\sum_\beta
|\langle E_\beta| \psi_0\rangle|^2 \,
\delta(E - E_\beta)
\end{equation}
is the spectral measure associated with the reference state $|\psi_0\rangle$.

In the thermodynamic limit, the spectrum becomes dense and we assume that the
spectral measure converges to an absolutely continuous form
$d\mu_0(E)=\rho_0(E)\,dE$, with no singular component. The Krylov resolvent
amplitudes then admit the integral representation
\begin{equation}
\psi_n(\xi)
= \int \frac{Q_n(E)}{E - \xi}\, \rho_0(E)\, dE.
\end{equation}

Under these assumptions, the functions $Q_n(E)$ form a family of orthonormal
polynomials with respect to the measure $\rho_0(E)dE$,
\begin{equation}
\int Q_n(E)\,Q_m(E)\,\rho_0(E)\,dE=\delta_{nm},
\end{equation}
and satisfy the three-term recurrence relation
\begin{equation}
b_{n+1} Q_{n+1}(E)
= (E-a_n)Q_n(E)-b_n Q_{n-1}(E),
\label{eq:recurrence}
\end{equation}
with $Q_{-1}(E)\equiv0$, $Q_0(E)\equiv1$, and $b_0=0$. This recurrence defines a
Jacobi operator whose spectral properties encode the structure of Krylov space.

For definiteness we assume that the spectral density has compact support
$\mathrm{supp}(\rho_0)=[E_{\min},E_{\max}]$ and write
\begin{equation}
\psi_n(\xi)
=
\int_{E_{\min}}^{E_{\max}}
\frac{Q_n(E)}{E-\xi}\,\rho_0(E)\,dE.
\label{eq:psidef}
\end{equation}
The behavior of $\psi_n(\xi)$ depends qualitatively on the position of $\xi$
relative to the spectral support.

Throughout the following analysis, we assume that the spectral measure is
absolutely continuous in a neighborhood of $\xi$, with a sufficiently smooth
density $\rho_0(E)$, and that the Lanczos coefficients admit well-defined
asymptotic limits. Under these assumptions, classical results from the theory of
orthogonal polynomials apply.

\subsection*{Case I: $|\xi|>E_{\max}$ (Outside the Spectrum)}

To analyze the behavior of Krylov resolvent amplitudes outside the spectral support, 
we first define the Stieltjes transform (or resolvent) associated with the spectral measure $\mu$ of the initial state:
\begin{equation}\label{eq:stieltjesdef}
R(\xi) = \int_{E_{\min}}^{E_{\max}} \frac{\rho(E)}{E-\xi} \, dE\,.
\end{equation}
For real $\xi$ lying outside the interval $[E_{\min}, E_{\max}]$, $R(\xi)$ is real-analytic, and we have
\begin{equation}
\psi_0(\xi) = R(\xi).
\end{equation}

The Krylov resolvent amplitudes can be expressed in terms of orthogonal polynomials $Q_n(\xi)$ and the associated polynomials of the second kind $P_{n-1}(\xi)$ \cite{Chihara1978}
\begin{equation}\label{eq:psiRQ}
\psi_n(\xi) = R(\xi) Q_n(\xi) - P_{n-1}(\xi),
\end{equation}
where $P_{-1}(\xi) \equiv 0$, $P_0(\xi) \equiv 1$, and for $n\ge1$,
\begin{equation}\label{eq:precurrence}
b_{n+1} P_n(\xi) = (\xi - a_n) P_{n-1}(\xi) - b_n P_{n-2}(\xi),
\end{equation}
with $\{a_n,b_n\}$ being the same Lanczos coefficients defining the tridiagonal Jacobi matrix corresponding to the Krylov space.

Assume that the Lanczos coefficients converge,
\begin{equation}
\lim_{n\to\infty} a_n = a_\infty,
\qquad
\lim_{n\to\infty} b_n = b_\infty > 0.
\end{equation}
In this case, 
the classical results from the theory of orthogonal polynomials
\cite{Szegoe,Deift,SimonOPRL}
then imply that for $\xi$ outside this interval the polynomials exhibit
exponential behavior
\begin{equation}
Q_n(\xi) \sim A(\xi)\, e^{n \kappa(\xi)} + B(\xi)\, e^{-n \kappa(\xi)}, 
\qquad \kappa(\xi) > 0.
\end{equation}
The square-summable solution corresponds to the decaying term, so that the Krylov resolvent amplitudes satisfy
\begin{equation}
|\psi_n(\xi)| \sim C(\xi)\, e^{-n \kappa(\xi)}.
\end{equation}

Consequently, the normalization sum converges:
\begin{equation}
\sum_{n=1}^\infty |\psi_n(\xi)|^2 \sim \frac{e^{-2\kappa(\xi)}}{1 - e^{-2\kappa(\xi)}} < \infty,
\end{equation}
and the corresponding probability distribution is exponentially localized.
The Krylov distribution, which measures the average spread of spectral weight along the Krylov chain, thus converges to a finite value in the thermodynamic limit $N \to \infty$:
\begin{equation}
\mathcal{D}(\xi) = \sum_{n=1}^N n\, P_n(\xi) \;\;\xrightarrow{N\to\infty}\;\; 
\frac{1}{1 - e^{-2 \kappa(\xi)}} < \infty.
\end{equation}

This analysis shows that, for energies outside the spectral support, the Krylov resolvent amplitudes decay exponentially along the chain, resulting in a Krylov distribution that is sharply localized near the first few layers. Physically, this reflects that inverse-energy processes are dominated by contributions from states near the spectral edge, with high-$n$ Krylov components effectively suppressed.

\subsection*{Case II: $\xi\in(E_{\min},E_{\max})$ (Inside the spectrum)}

When $\xi$ lies within the spectral support, the integral
\eqref{eq:psidef} must be interpreted in the principal-value sense,
\begin{equation}
\psi_n(\xi)
=
\mathrm{P.V.}\!\int_{E_{\min}}^{E_{\max}}
\frac{Q_n(E)}{E-\xi}\,\rho_0(E)\,dE.
\end{equation}
The large-$n$ behavior of $Q_n(E)$ in the interior of the spectrum is oscillatory.
Away from spectral edges, the leading Plancherel--Rotach asymptotics read~\cite{Szegoe,Deift,SimonOPRL}
\begin{equation}
Q_n(\xi)
\sim
\sqrt{\frac{2}{\pi\rho_0(\xi)}}
\frac{\sin\!\bigl(n\theta(\xi)+\delta(\xi)\bigr)}
     {\sqrt{\sin\theta(\xi)}},
\end{equation}
where
\begin{equation}
\theta(\xi)=\pi\int_{E_{\min}}^\xi \rho_0(E)\,dE.
\end{equation}

Applying the Sokhotski--Plemelj formula, one finds that the imaginary part of
$\psi_n(\xi\pm i0)$ is proportional to $\rho_0(\xi)Q_n(\xi)$, while the
principal-value contribution remains bounded due to oscillatory cancellations.
Consequently, the dominant large-$n$ contribution to $|\psi_n(\xi)|^2$ scales as
\begin{equation}
|\psi_n(\xi)|^2 \sim \pi^2\rho_0(\xi)^2 |Q_n(\xi)|^2.
\end{equation}
Averaging over oscillations yields an $O(1)$ contribution per Krylov layer,
implying
\begin{equation}
\sum_{n=0}^{N-1}|\psi_n(\xi)|^2 \sim C(\xi)\,N,
\end{equation}
and hence $P_n(\xi)\sim 1/N$. The Krylov distribution therefore grows
extensively,
\begin{equation}
\mathcal{D}(\xi)\sim \frac{N}{2},
\qquad N\to\infty.
\end{equation}

Physically, this reflects the fact that for spectral parameters $\xi$ inside the
spectrum the resolvent-dressed state remains delocalized along Krylov space in an
averaged sense. In contrast to the exponentially localized behavior outside the
spectrum, the inverse-energy weight spreads approximately uniformly across Krylov
layers, signaling the absence of Krylov-space localization.

\subsection*{Case III: Spectral Edges and Critical Points}

In the preceding analysis, we focused on resolvent parameters $\xi$ lying either deep in the bulk of a continuous spectrum or at a finite distance outside it. In these regimes, the asymptotic behavior of the Krylov resolvent amplitudes and the associated distribution $\mathcal{D}(\xi)$ is relatively straightforward. At spectral edges and interacting quantum critical points, however, the density of states $\rho_0(E)$ develops singular behavior.
As a result, the asymptotics of both the orthogonal polynomials and the Krylov
resolvent amplitudes are qualitatively modified, leading to distinct scaling
regimes for the Krylov distribution (see Appendix~\ref{app:asymptotics} for more details).

For a broad class of physically relevant systems, including free fermions, tight-binding models, and random matrix ensembles, the density of states near a regular band edge $E_*$ exhibits a square-root behavior,
\begin{equation}
\rho_0(E) \sim C\,(E-E_\ast)^{1/2},
\qquad E \gtrsim E_\ast,
\end{equation}
with an analogous behavior near upper band edges.
This scaling follows generically from quadratic extrema of the underlying
dispersion relation and is largely insensitive to microscopic details \cite{MehtaRandomMatrices}.

The square-root edge represents a turning-point problem for the associated Jacobi operator. As a result, the orthogonal polynomials exhibit universal Airy-type asymptotics \cite{SimonOPRL}
\begin{equation}
Q_n(E_\ast + x n^{-2/3}) \sim n^{-1/3} \,\mathrm{Ai}(c\,x),
\end{equation}
where $\mathrm{Ai}(x)$ is the Airy function and $c$ is a model-dependent constant. Substituting these asymptotics into the spectral representation of the resolvent, one finds algebraic decay of the Krylov amplitudes
\begin{equation}
|\psi_n(\xi)|^2 \sim n^{-4/3},
\end{equation}
up to oscillatory and logarithmic corrections. Consequently, the Krylov distribution grows sublinearly with the Krylov cutoff $N$,
\begin{equation}
\mathcal{D}(\xi) \sim N^{2/3}.
\end{equation}
This scaling reflects the critical slowing-down of resolvent propagation at a regular spectral edge and is a generic for smooth band edges
that is the  manifestation of Airy-type edge physics.

More generally, one may encounter edges where the density of states behaves as a power law
\begin{equation}
\rho_0(E) \sim (E - E_\ast)^\alpha, \qquad \alpha>-1\,,
\end{equation}
but does not correspond to a turning point of the Jacobi operator. In these non-generic edges, stationary-phase and integration-by-parts methods remain valid, leading to
\begin{equation}
|\psi_n(\xi)|^2 \sim n^{-2(1+\alpha)},
\end{equation}
and thus to the Krylov distribution
\begin{equation}
\mathcal{D}(\xi) \sim
\begin{cases}
\text{constant}, & \alpha >0\,,\\
\ln N, & \alpha =0\,,\\
N^{-2\alpha}, & -\frac{1}{2}<\alpha < 0,\\
N/\ln N, & \alpha =- \frac{1}{2},\\
N, & \alpha <- \frac{1}{2}.
\end{cases}
\end{equation}

The case of $\alpha=0$ has an interesting interpretation. Indeed,
at interacting quantum critical points, the low-energy density of states may
approach a constant or exhibit other singular behavior reflecting scale
invariance~\cite{Sachdev:2011},
\begin{equation}
\rho_0(E) \sim E^{\frac{1}{z}-1},
\end{equation}
where $z$ is the dynamical critical exponent.
In one-dimensional relativistic critical systems ($z=1$),
$\rho_0(E)\sim \mathrm{const}$, leading to algebraic decay of the Krylov
amplitudes,
\begin{equation}
|\psi_n(\xi)|^2 \sim \frac{1}{n^2},
\qquad
\mathcal{D}(\xi) \sim \log N,
\end{equation}
so that logarithmic growth of the Krylov distribution arises naturally in
gapless critical systems \cite{Sachdev:2011}.

To conclude this section we have seen that  the thermodynamic behavior of the Krylov distribution falls into three universal classes. Outside the spectrum or near isolated eigenvalues, the resolvent is localized and $\mathcal{D}(\xi)=O(1)$. In the bulk of a continuous spectrum, the Krylov amplitudes are delocalized and $\mathcal{D}(\xi)\sim N/2$. Near spectral edges and quantum critical points, the distribution grows sublinearly or logarithmically depending on the universality class. These results demonstrate that $\mathcal{D}(\xi)$ provides a sharp and quantitative probe of spectral singularities, universality classes, and the large-scale structure of Krylov space.


\section{Explicit Exactly Solvable Models}
\label{sec:Three-Model}

In this section we present three explicit, exactly solvable models for which the
Krylov resolvent amplitudes $\psi_n(\xi)$ can be computed analytically. These
models serve complementary purposes and illustrate the full range of behaviors
discussed in the general framework. First, the model with constant Lanczos
coefficients provides a benchmark for systems with continuous, bounded spectra
and demonstrates how resolvent states explore Krylov space uniformly. Second, the
quadratic (displaced oscillator) Hamiltonian exhibits linearly growing diagonal
coefficients with square-root hopping, leading to a discrete, unbounded spectrum
with strong localization in Krylov space except at resonant energies. Third, the
$SU(1,1)$ chain features linear growth of the off-diagonal coefficients
$b_n \sim \alpha n$, modeling the asymptotic behavior of maximally chaotic
systems like the SYK model and exhibiting continuous spectra with power-law
tails in Krylov space. 

Together, these examples make transparent the asymptotic
behavior of the Krylov distribution $\mathcal{D}(\xi)$ in different spectral
regimes and illuminate the relationship between spectral properties, Lanczos
coefficient growth, and Krylov-space geometry.

\subsection{Model with Constant Lanczos Coefficients}

We now illustrate the general asymptotic framework developed above through a
simple but instructive exactly solvable example. This model provides a concrete
realization of the three universal regimes—outside the spectrum, inside the
bulk, and at a spectral edge—and highlights the distinction between generic and
nongeneric edge behavior.

We consider the case of constant Lanczos coefficients,
\begin{equation}
a_n = 0, \qquad b_n = b, \qquad n \ge 1,
\end{equation}
defining a Jacobi operator with uniform hopping and vanishing on-site potential. 
The Krylov recursion then becomes a translation-invariant second-order difference equation, allowing for a fully explicit spectral analysis.

A canonical realization is a single particle hopping on an infinite one-dimensional tight-binding lattice,
\begin{equation}
H = -b \sum_{n \in \mathbb{Z}} \left(|n\rangle\langle n+1| + |n+1\rangle\langle n|\right),
\end{equation}
with initial state $|\psi_0\rangle = |0\rangle$. The Lanczos algorithm generates a Krylov basis of symmetric superpositions at increasing distance from the origin, yielding exactly $b_n = b$. A constant shift $a_n = a$ rigidly shifts the spectrum but does not alter the scaling of Krylov quantities with $N$.

For $n \ge 1$, the resolvent amplitudes satisfy
\begin{equation}
b\,\psi_{n+1} - \xi\,\psi_n + b\,\psi_{n-1} = 0,
\end{equation}
with characteristic equation
\begin{equation}
b r^2 - \xi r + b = 0, \qquad r_\pm = \frac{\xi \pm \sqrt{\xi^2 - 4b^2}}{2b}.
\end{equation}

\paragraph{Outside the spectrum ($|\xi|>2b$).}
The roots $r_\pm$ are real and reciprocal. Normalizability selects the decaying root $|r|<1$, giving
\begin{equation}
\psi_n(\xi) \sim r^n, \qquad |r| < 1.
\end{equation}
The Krylov distribution is geometric $P_n = (1 - |r|^2) |r|^{2n},$
\begin{equation}
\mathcal{D}(\xi) = \frac{|r|^2}{1 - |r|^2} = \frac12\left(\frac{|\xi|}{\sqrt{\xi^2 - 4b^2}} - 1\right),
\end{equation}
which is finite and independent of $N$, exemplifying Krylov localization outside the spectral band.

\paragraph{Inside the spectrum ($|\xi|<2b$).}
Writing $\xi = 2b \cos k$, the roots are $r_\pm = e^{\pm i k}$ and the resolvent amplitudes are oscillatory with constant envelope:
\begin{equation}
\psi_n(\xi) \sim \frac{1}{\sqrt{4b^2 - \xi^2}} \, e^{i k n}, \qquad |\psi_n|^2 \sim \frac{1}{4b^2 - \xi^2}.
\end{equation}
Normalization over a cutoff $N$ yields $P_n \sim 1/N$, giving
\begin{equation}
\mathcal{D}_N(\xi) \sim \frac{N}{2}, \quad N \gg 1,
\end{equation}
signaling Krylov delocalization.

\paragraph{Spectral edge ($\xi \to 2b$).}
Near the band edge, the plane-wave approximation breaks down.\footnote{Using the parametrization $\xi = 2b \cos k$, the edge corresponds to $k = 0$. The group velocity $v_g = d\xi/dk = -2b \sin k$ then vanishes, signaling a turning-point regime. In this region, the Krylov amplitudes are no longer well approximated by plane waves and must instead be described by Airy functions.}  
In this regime, the envelope of the amplitudes varies slowly with $n$. Introducing the scaled variable
\begin{equation}
x \sim n^{2/3} (2b - \xi),
\end{equation}
the resolvent amplitudes satisfy the Airy equation (see Appendix \ref{app:asymptotics})
\begin{equation}
\psi_n(\xi) \sim n^{-1/3} \mathrm{Ai}\!\Big(b^{-1/3} n^{2/3} (2b-\xi)\Big),
\end{equation}
leading to a subextensive Krylov distribution
\begin{equation}
\mathcal{D}_N(\xi) \sim N^{2/3}, \qquad \xi \approx 2b.
\end{equation}
This is the universal edge scaling, arising from the turning-point physics of Jacobi operators with asymptotically constant Lanczos coefficients.

These results connect directly to the critical one-dimensional transverse-field
Ising model, which maps to free fermions with Hamiltonian
\begin{equation}
H = \sum_k \epsilon_k \left(\gamma_k^\dagger \gamma_k - \frac{1}{2}\right),
\qquad
\epsilon_k = 4J \left|\sin\frac{k}{2}\right|.
\end{equation}
Deep inside the gapless spectrum, the asymptotic Lanczos coefficients saturate to
$b_n \to 2J$, and the Krylov resolvent amplitudes remain of order unity across the
chain. Consequently, the Krylov distribution grows linearly, $\mathcal{D}\sim N$,
signaling ballistic spreading in Krylov space. By contrast, in gapped phases the
spectral gap induces exponential decay of amplitudes beyond a finite
localization length $\xi \sim 2J/\Delta$, causing $\mathcal{D}$ to saturate.

The constant-$b_n$ model thus captures the essential distinction between gapped
and gapless phases in the simplest possible setting, while simultaneously
highlighting the difference between fine-tuned and universal spectral-edge
behavior.

\subsection{Quadratic Hamiltonian: Particle in the Heisenberg--Weyl Group}

We now examine a qualitatively different exactly solvable example corresponding
to a Krylov chain with linearly growing diagonal coefficients and square-root
hopping:
\begin{equation}
a_n = \omega n, \qquad b_n = \lambda \sqrt{n}, \quad n \ge 1,
\end{equation}
which arises from the quadratic Hamiltonian
\begin{equation}
H = \omega a^\dagger a + \lambda (a + a^\dagger).
\end{equation}
This integrable model describes a displaced harmonic oscillator with a discrete,
unbounded spectrum
\begin{equation}
E_m = \omega m - \gamma^2 \omega, \qquad m = 0,1,2,\dots,
\end{equation}
where $\gamma = \lambda/\omega$.

The exact time-evolved Krylov amplitudes are known analytically
\cite{Balasubramanian:2022tpr}:
\begin{equation}
\phi_n(t) = e^{\alpha(t)} \frac{z(t)^n}{\sqrt{n!}},
\end{equation}
with
\begin{align}
z(t) &= -\gamma\bigl(1 - e^{-i\omega t}\bigr),\\
\alpha(t) &= \gamma^2\bigl(i\omega t - 1 + e^{-i\omega t}\bigr).
\end{align}

The resolvent amplitudes $\psi_n(\xi)$ follow from the Laplace transform and can
be evaluated exactly (see Appendix~\ref{app:details} for derivation):
\begin{equation}\label{eq:psi_quad_exact}
\psi_n(\ell) =
\frac{(-1)^{n+1} \gamma^{n}}{\omega e^{\gamma^2}}
\frac{\Gamma(-\ell)\sqrt{n!}}{\Gamma(n+1-\ell)}
\, {}_1F_1(-\ell,n+1-\ell,\gamma^2),
\end{equation}
where ${}_1F_1$ denotes the confluent hypergeometric function and we have set
$\xi = E_\ell$ with $\ell$ treated as a real parameter.

The explicit expression \eqref{eq:psi_quad_exact} makes the resonant structure of
the resolvent fully transparent. As the resolvent parameter $\xi$ approaches a
physical eigenvalue $E_m$, the Gamma function $\Gamma(-\ell)$ develops a pole,
leading to a pronounced enhancement of the Krylov resolvent amplitudes.
Consequently, the Krylov distribution exhibits sharp resonances at these values
of $\xi$. Away from resonance, the amplitudes decay rapidly with Krylov index
$n$, resulting in strong localization of the resolvent-dressed state near the
beginning of the Krylov chain. The exact resonance values, derived analytically
in Appendix~\ref{app:details}, are given by
\begin{equation}
\mathcal{D}(E_m) = m + \gamma^2, 
\qquad m = 0,1,2,\dots\,.
\end{equation}

\begin{figure}[h]
\centering
\includegraphics[width=0.46\textwidth]{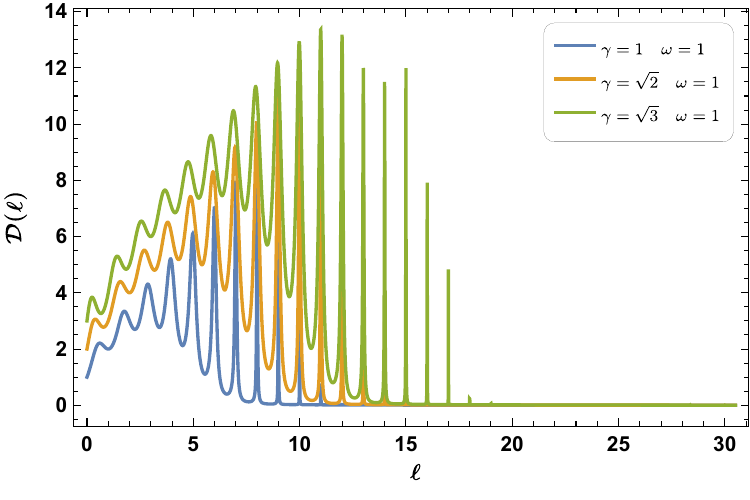}
\caption{Krylov distribution $\mathcal{D}(\ell)$ for the quadratic Hamiltonian,
computed numerically from the exact resolvent amplitudes
Eq.~\eqref{eq:psi_quad_exact} with a Krylov cutoff $N=25$. Sharp peaks appear at
the discrete eigenvalues $E_m$, reflecting the pole structure of the resolvent.
At resonance, the Krylov distribution spreads over many layers, with peak
heights growing linearly with $m$ according to
$\mathcal{D}(E_m) = m + \gamma^2$. In contrast, off-resonant values of $\ell$
exhibit strong Krylov-space localization, with $\mathcal{D}(\ell)$ remaining of
order unity.}
\label{D-Har}
\end{figure}

Numerical evaluation of the exact resolvent amplitudes
Eq.~\eqref{eq:psi_quad_exact} with a finite Krylov cutoff $N=25$ clearly
illustrates this structure, as shown in Fig.~\ref{D-Har}. The Krylov distribution
$\mathcal{D}(\ell)$ displays sharp, isolated peaks at the discrete eigenvalues
$E_m$ of the quadratic Hamiltonian. At these resonances, the distribution spreads
over many Krylov layers, and the peak heights follow
$\mathcal{D}(E_m) \simeq m + \gamma^2$, in excellent agreement with the analytic prediction. In contrast, away from the
eigenvalues the resolvent amplitudes are strongly suppressed, yielding a
localized Krylov distribution with $\mathcal{D}(\ell) \lesssim \mathcal{O}(1)$.

Figure~\ref{D-Har} highlights several essential features of the Krylov resolvent
structure. First, the discrete pole structure of the resolvent manifests as
isolated peaks in $\mathcal{D}(\ell)$ at each eigenvalue. Second, two sharply
distinct regimes emerge: a resonant regime, in which the Krylov distribution is
broad and explores a large number of Krylov layers, and an off-resonant regime,
in which the distribution remains narrowly concentrated near the lowest Krylov
indices. Finally, the linear growth of the peak heights with $m$ directly
reflects the analytic result $\mathcal{D}(E_m) = m + \gamma^2$, providing a
nontrivial consistency check of the exact solution.

\begin{figure}[h]
\centering
\includegraphics[width=0.46\textwidth]{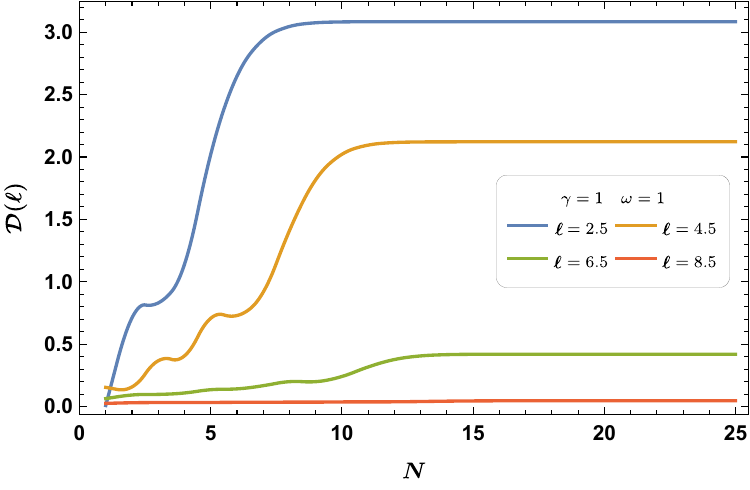}
\caption{Krylov distribution $\mathcal{D}(\ell)$ as a function of the Krylov cutoff
$N$ for the quadratic Hamiltonian, computed numerically from
Eq.~\eqref{eq:psi_quad_exact} for several fixed values of $\ell$. The distribution
grows approximately linearly at small $N$ and saturates at larger $N$,
reflecting the finite support of the resolvent-dressed state in Krylov space.
Oscillatory features originate from near-resonant enhancement when $\xi$
approaches an eigenvalue; choosing half-integer values of $\ell$ suppresses these
oscillations and yields smoother scaling behavior.}
\label{D-HarDN}
\end{figure}

To further examine the dependence of the Krylov distribution on the cutoff $N$,
we compute $\mathcal{D}(\ell)$ numerically as a function of $N$ for several fixed
values of the resolvent parameter $\ell$. The results are shown in
Fig.~\ref{D-HarDN}. For generic values of $\ell$, the Krylov distribution grows
approximately linearly with $N$ at small $N$ before saturating, reflecting the
finite extent of the resolvent-dressed state in Krylov space. Superimposed
oscillations arise from near-resonant enhancement when $\xi$ approaches an
eigenvalue. To suppress these oscillations and expose the underlying scaling
behavior more clearly, we also consider half-integer values of $\ell$, which avoid
exact resonances.

\subsection{$SU(1,1)$ Chain: Continuous Spectrum and Power-Law Tails}

As the third example we study a paradigmatic model in Krylov space defined by
Lanczos coefficients
\begin{equation}
a_n = 0, \qquad b_n = \alpha\sqrt{n(n+h-1)}, \quad h,\alpha>0.
\end{equation}
This model possesses an exact $SU(1,1)$ symmetry, with Hamiltonian
$H = \alpha(L_+ + L_-)$, where $L_\pm$ are the raising and lowering operators of
the discrete series representation of $SU(1,1)$ \cite{Balasubramanian:2022tpr}.
Physically, it describes quantum dynamics governed by hyperbolic rather than
oscillatory motion, making it a prototype for understanding complexity growth in
chaotic systems with continuous spectra \cite{Parker:2018yvk}.

This model captures the asymptotic linear growth of Lanczos coefficients characteristic of a broad class of systems with continuous spectra, including maximally chaotic models such as Sachdev-Ye-Kitaev (SYK) model \cite{Sachdev:1993bu,Kitaev:2015,Kitaev:2017aw}.
This suggests the $SU(1,1)$ chain captures essential features of quantum chaos:
unbounded operator growth \cite{Parker:2018yvk,Rabinovici:2020ryf}, continuous
spectral characteristics, and power-law decay of correlations in Krylov space.
While the SYK model exhibits $b_n \sim \pi \mathcal{J} n$ at large $n$
\cite{Balasubramanian:2022tpr,Rabinovici:2022beu}, with $\mathcal{J}$ the
coupling strength, this model provides an analytically tractable realization of
this linear growth paradigm.

The model is exactly solvable. The time-evolved amplitudes are given in closed
form as \cite{Parker:2018yvk,Balasubramanian:2022tpr}
\begin{equation}
\phi_n(t) = \sqrt{\frac{(h)_n}{n!}}\; \frac{\tanh^n(\alpha t)}{\cosh^{h}(\alpha t)},
\end{equation}
where $(h)_n = h(h+1)\cdots(h+n-1)$ is the Pochhammer symbol. These satisfy
$\sum_{n=0}^\infty |\phi_n(t)|^2 = 1$ for all $t$. Their Laplace transforms, or
Krylov resolvent amplitudes, are (see Appendix~\ref{sec:su11_model} for details)
\begin{align}\label{exact-SU11}
\psi_n(\xi) =& -\frac{i}{\alpha}
\frac{\Gamma(a)\sqrt{(h)_n n!}}{\Gamma(n+a+1)}\nonumber\\ &
\times {}_2F_1\!\left(1-b,\; n+1;\; n+a+1;\; -1\right),
\end{align}
with $a = i\xi/(2\alpha) + h/2$ and $b = -i\xi/(2\alpha) + h/2$.

The analytic structure of $\psi_n(\xi)$ reveals poles at
$\xi_m = i\alpha(2m+h)$ on the imaginary axis, indicating the absence of discrete
real eigenvalues. This contrasts with the quadratic model, where poles occur at
real energies $E_m$. The $SU(1,1)$ Hamiltonian therefore exhibits a continuous
spectrum, a hallmark of non-compact algebraic structures.

For real $\xi$, the asymptotic behavior of the Krylov resolvent amplitudes is
universal:
\begin{equation}
|\psi_n(\xi)|^2 \sim \frac{K(\xi,h)}{n} \quad \text{as } n \to \infty.
\end{equation}
The $SU(1,1)$ chain exhibits hyperbolic dynamics fundamentally distinct from the oscillatory motion of compact groups. The initial state $|0\rangle$, corresponding to the lowest weight state of the non-compact algebra, spreads indefinitely under time evolution without revivals, leading to unbounded growth of Krylov complexity. This behavior is mathematically encoded in the power-law tail $|\psi_n|^2 \sim n^{-1}$, which signals a continuous spectral measure in Krylov space analogous to gapless systems. The continuous spectrum arises from the non-compact nature of $SU(1,1)$, contrasting sharply with the discrete spectrum of the quadratic model governed by the compact Heisenberg–Weyl algebra. 

This $n^{-1}$ power-law tail has profound physical consequences. The distribution is not strictly normalizable in the infinite-$N$ limit, as $\sum_n |\psi_n|^2$ diverges logarithmically. Consequently, the Krylov distribution exhibits the characteristic large-$N$ scaling:
\begin{equation}
\mathcal{D}_N(\xi) \sim \frac{N}{\ln N}.
\label{eq:su11_scaling}
\end{equation}
This behavior stands in stark contrast to the quadratic model, where at discrete resonances the distribution saturates to finite values $\mathcal{D}(E_m) = m + \gamma^2$. The divergence of $\mathcal{D}_N(\xi)$ with $N$ reflects the continuous nature of the spectrum in the $SU(1,1)$ model, where resolvent states remain delocalized across arbitrarily high Krylov indices.

\begin{figure}[h]
\centering
\includegraphics[width=0.4\textwidth]{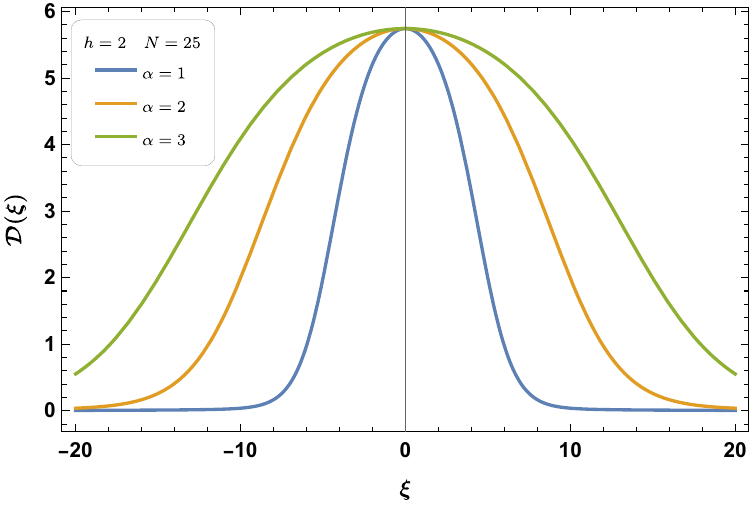}
\caption{
Krylov distribution $\mathcal{D}(\xi)$ for the $SU(1,1)$ chain with $h=2$ and Krylov cutoff $N=25$, plotted as a function of the spectral parameter $\xi$. 
The curves correspond to different values of the representation parameter $\alpha$: $\alpha=1$ (blue), $\alpha=2$ (red), and $\alpha=3$ (green), and are computed numerically from the exact resolvent amplitudes in Eq.~\eqref{exact-SU11}. 
The smooth, non-resonant dependence on $\xi$ reflects the absolutely continuous spectrum associated with the non-compact $SU(1,1)$ algebra and the absence of Krylov-space localization.
}
\label{D-SU11}
\end{figure}

The resulting Krylov distribution $\mathcal{D}(\xi)$, computed numerically from the exact resolvent amplitudes, is shown in Fig.~\ref{D-SU11} as a function of the spectral parameter $\xi$ for a fixed Krylov cutoff $N=25$ and $h=2$, and for several values of the representation parameter $\alpha$. This figure illustrates how the detailed shape of $\mathcal{D}(\xi)$ depends on the choice of $SU(1,1)$ representation while remaining smooth and non-singular across the spectrum.

The curves display a smooth, Gaussian-like maximum centered at $\xi=0$ with identical height for all $\alpha$, followed by gradual decay to zero at larger $|\xi|$. The maximum height is universal because at $\xi=0$ the parameter $\alpha$ enters $\psi_n(0)$ only as an overall factor $1/\alpha$, which cancels in the normalized distribution $P_n = |\psi_n|^2/\sum_m |\psi_m|^2$, leaving $\mathcal{D}(0) = \sum_n n P_n(0)$ independent of $\alpha$. The decay is faster for smaller $\alpha$, indicating that $\alpha$ controls the spectral width in Krylov space: a smaller $\alpha$ corresponds to a narrower effective bandwidth, causing the resolvent-dressed state to localize more sharply near $\xi=0$. Physically, $\alpha$ sets the energy scale of the $SU(1,1)$ Hamiltonian $H=\alpha(L_+ + L_-)$; a smaller $\alpha$ reduces the rate of Krylov hopping, thereby narrowing the spectral support of the resolvent. This behavior contrasts with the quadratic model, where sharp peaks occur at discrete eigenenergies—here, the smooth, Gaussian-like maximum at $\xi=0$ reflects the continuous, gapless nature of the spectrum, with $\alpha$ acting as a broadening parameter that sets the scale of spectral decay away from the origin in Krylov space.

\begin{figure}[h]
\centering
\includegraphics[width=0.4\textwidth]{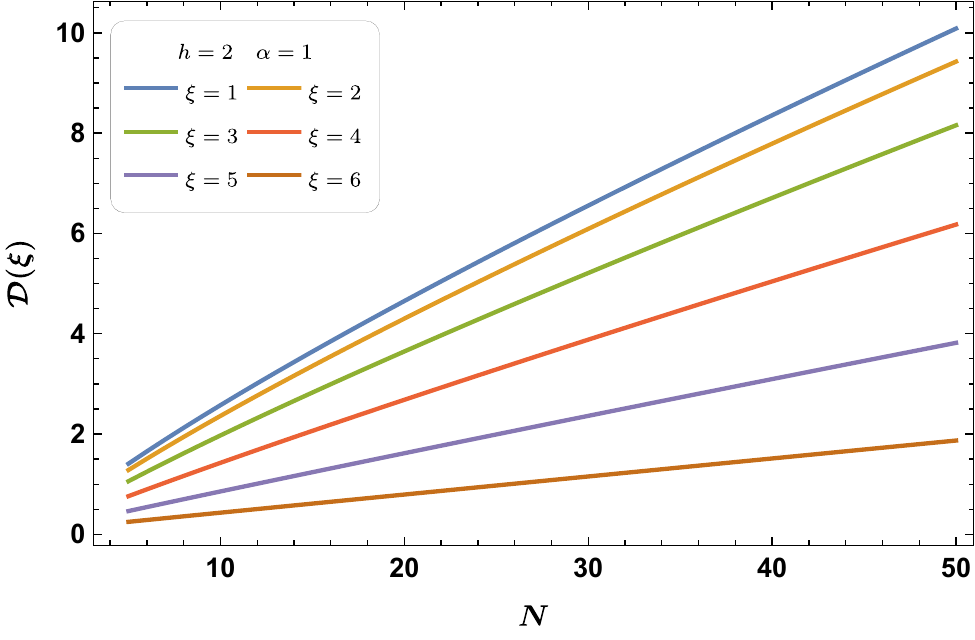}
\caption{
Krylov distribution $\mathcal{D}(\xi)$ for the $SU(1,1)$ chain with $h=2$ and $\alpha=1$, plotted as a function of the Krylov cutoff $N$ for several fixed values of the spectral parameter $\xi$ (indicated in the inset).
The growth of $\mathcal{D}(\xi)$ is sublinear and consistent with the asymptotic $N/\ln N$ scaling expected for this model, illustrating the slow but delocalized exploration of Krylov space induced by the underlying non-compact algebraic structure.
}
\label{D-SU11DN}
\end{figure}

Complementarily, Fig.~\ref{D-SU11DN} displays $\mathcal{D}(\xi)$ as a function of the Krylov cutoff $N$ for fixed $\alpha=1$ and several representative values of $\xi$. This presentation highlights the scaling of the Krylov distribution with system size and allows for direct comparison with the expected asymptotic behavior. In particular, the numerical data clearly demonstrate sublinear growth consistent with the analytically predicted $N/\log N$ scaling characteristic of the $SU(1,1)$ chain.

The $SU(1,1)$ chain serves as a benchmark for understanding how continuous spectra manifest in Krylov space, exhibiting smooth, Gaussian-like distributions rather than discrete resonances. Its linear growth $b_n \sim \alpha n$ matches the asymptotic behavior of maximally chaotic systems like the SYK model, while remaining analytically tractable. The $n^{-1}$ tail and the universal maximum of $\mathcal{D}(\xi)$ at $\xi=0$—which is independent of $\alpha$ due to normalization—provide a testing ground for spectral diagnostics in gapless systems. Furthermore, $SU(1,1)$ symmetry appears in diverse physical contexts, including conformal field theories, black hole physics, and squeezed states in quantum optics, making this model broadly relevant across different domains while remaining a minimal prototype for understanding the relationships between algebraic structure, spectral properties, and Krylov-space distributions in systems with continuous spectra.

\subsection{Comparison of the Three Models}

These exactly solvable models collectively illustrate how spectral properties govern Krylov-space exploration, revealing the intimate relationship between Lanczos coefficient growth, spectral structure, and Krylov-space geometry.

The comparison reveals a clear hierarchy of behaviors: constant $b_n$ produces extensive spreading ($\mathcal{D} \sim N$); square-root-growing $b_n$ with a discrete spectrum results in localized amplitudes punctuated by resonant peaks; and linearly growing $b_n$ with a continuous spectrum generates power-law tails and logarithmically enhanced complexity growth. Together, these three models illustrate how spectral properties---bounded versus unbounded, discrete versus continuous---are encoded in Krylov space, manifesting as distinct scaling laws for the Krylov distribution $\mathcal{D}(\xi)$.

\section{Krylov Distribution for the Ising Model}
\label{sec:Ising}

In this section, we examine the Krylov distribution $\mathcal{D}(\xi)$ for a
spin-$\frac{1}{2}$ mixed-field Ising chain governed by the Hamiltonian
\begin{equation}
\label{eq:Ising}
H = -J \sum_{i=1}^{L-1} \sigma_i^z \sigma_{i+1}^z
    - \sum_{i=1}^{L} \left( g \sigma_i^x + h \sigma_i^z \right).
\end{equation}
Here $\sigma_i^{x,y,z}$ denote Pauli matrices acting on site $i$, and $J$, $g$, and
$h$ are real coupling parameters. Throughout this section, we set $J=1$ by
rescaling energies.

The qualitative properties of the model depend sensitively on the transverse and
longitudinal fields $g$ and $h$. For $h=0$, the model is integrable and can be
mapped to free fermions via the Jordan--Wigner transformation. In this case, the
system undergoes a quantum phase transition at $g=1$, where the excitation gap
closes and the Lanczos coefficients approach constants at large Krylov index.
When both $g\neq0$ and $h\neq0$, integrability is broken and the system exhibits
quantum-chaotic spectral properties characterized by level repulsion and
spectral rigidity.

The purpose of this section is to explore how the Krylov distribution
$\mathcal{D}(\xi)$ reflects these distinct spectral structures and how its
behavior depends on the choice of reference state. Unlike time-dependent probes
such as Krylov complexity, the Krylov distribution is defined through the
resolvent and is therefore directly sensitive to spectral organization rather
than dynamical spreading. 

To probe the spectrum from different perspectives, we consider two classes of
initial states. A first class consists of site-factorized product states,
\begin{equation}
\label{eq:initial}
|\theta,\phi\rangle =
\prod_{i=1}^{L}
\left(
\cos\frac{\theta}{2}\, |Z+\rangle_i
+ e^{i\phi}\sin\frac{\theta}{2}\, |Z-\rangle_i
\right),
\end{equation}
where $|Z\pm\rangle_i$ are eigenstates of $\sigma_i^z$ with eigenvalues $\pm1$.
We restrict to homogeneous configurations with identical angles $(\theta,\phi)$
on each site. These states are maximally localized in real space and strongly
biased toward specific operator sectors, making them sensitive to fine spectral
features.

To probe more global spectral properties, we also consider a Gibbs-like
superposition of energy eigenstates,
\begin{equation}
\label{eq:TS}
|\psi_0\rangle =
\frac{1}{\sqrt{Z(\beta)}}
\sum_{j \in I} e^{-\frac{1}{2}\beta E_j}\, |E_j\rangle,
\qquad
Z(\beta)=\sum_{j \in I} e^{-\beta E_j},
\end{equation}
where $|E_j\rangle$ are eigenstates of $H$ and $I$ denotes a chosen subset of the
spectrum. By varying $I$, one may construct states localized in narrow energy
windows or broadly distributed across the spectrum. When $I$ includes all
eigenstates, the state samples the full Hilbert space with a weak energy bias
controlled by $\beta$. Since the Hamiltonian~\eqref{eq:Ising} possesses a
$\mathbb{Z}_2$ parity symmetry~\cite{Joel:2013}, the sum is restricted to a fixed
parity sector.

We focus on a chain of length $L=9$ and fix $g=-1.05$. Two representative regimes
are considered: the integrable case $h=0$ and the chaotic case $h=0.5$. For each
choice of parameters and initial state, we compute the Krylov distribution
$\mathcal{D}(\xi)$ as a function of the resolvent parameter $\xi$ by numerically
constructing the Krylov basis via the Lanczos algorithm and evaluating the
corresponding resolvent amplitudes. The Lanczos coefficients of this model have
been studied extensively in previous work
\cite{Scialchi:2023bmw,Trigueros:2021rwj,Espanol:2022cqr,Noh_2021,
Bhattacharya:2023zqt,Bhattacharya:2022gbz,Alishahiha:2024rwm}.

\begin{figure}[t]
\centering
\includegraphics[width=0.32\textwidth]{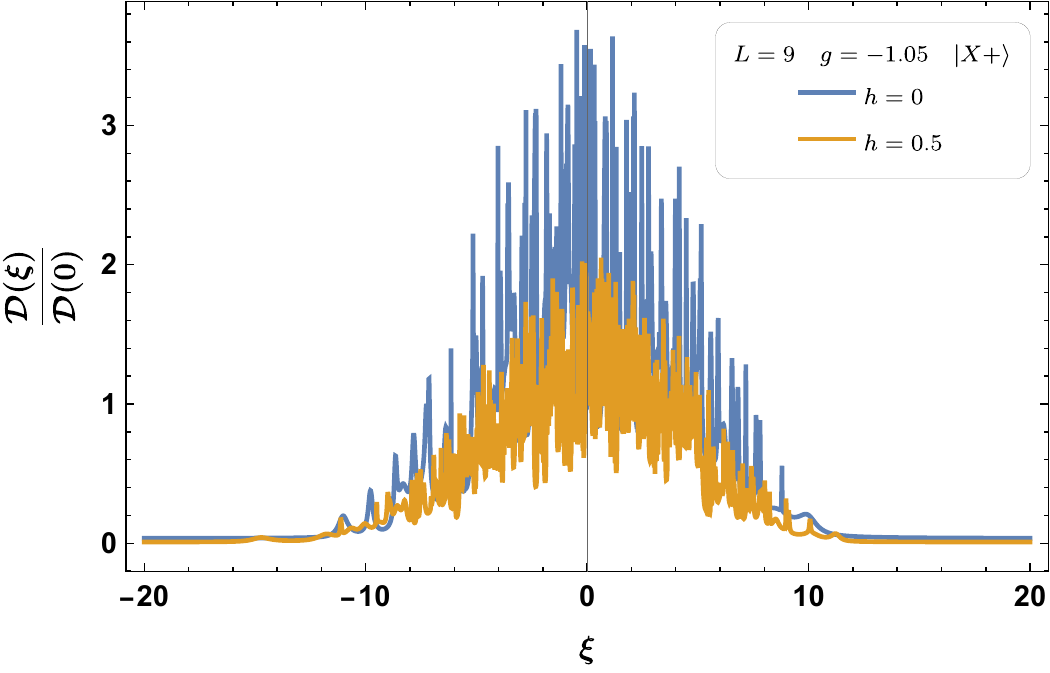}
\includegraphics[width=0.32\textwidth]{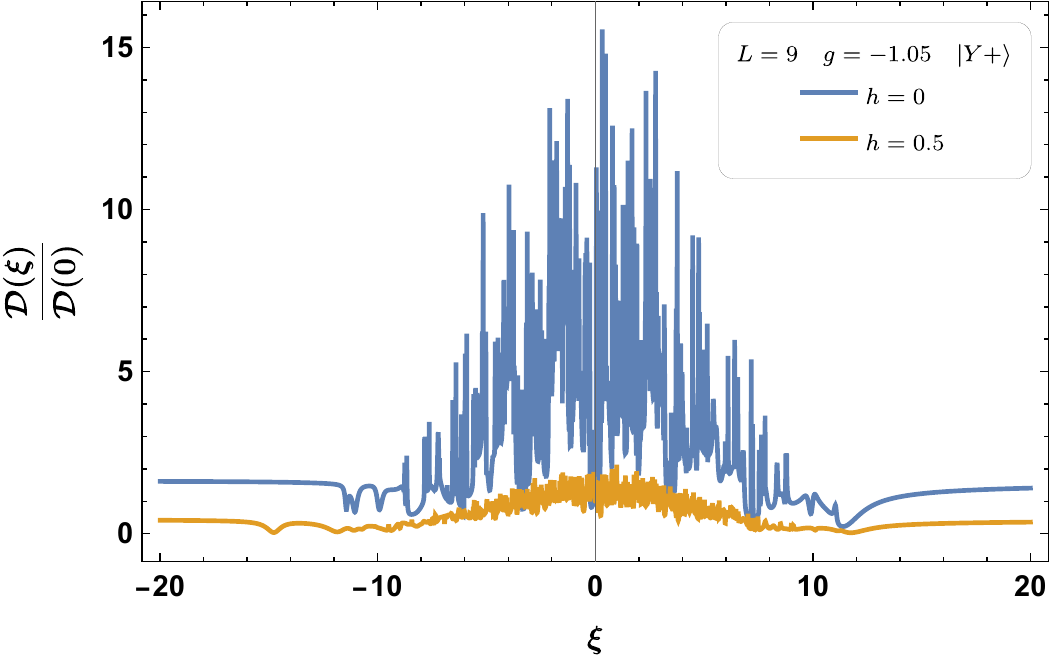}
\includegraphics[width=0.32\textwidth]{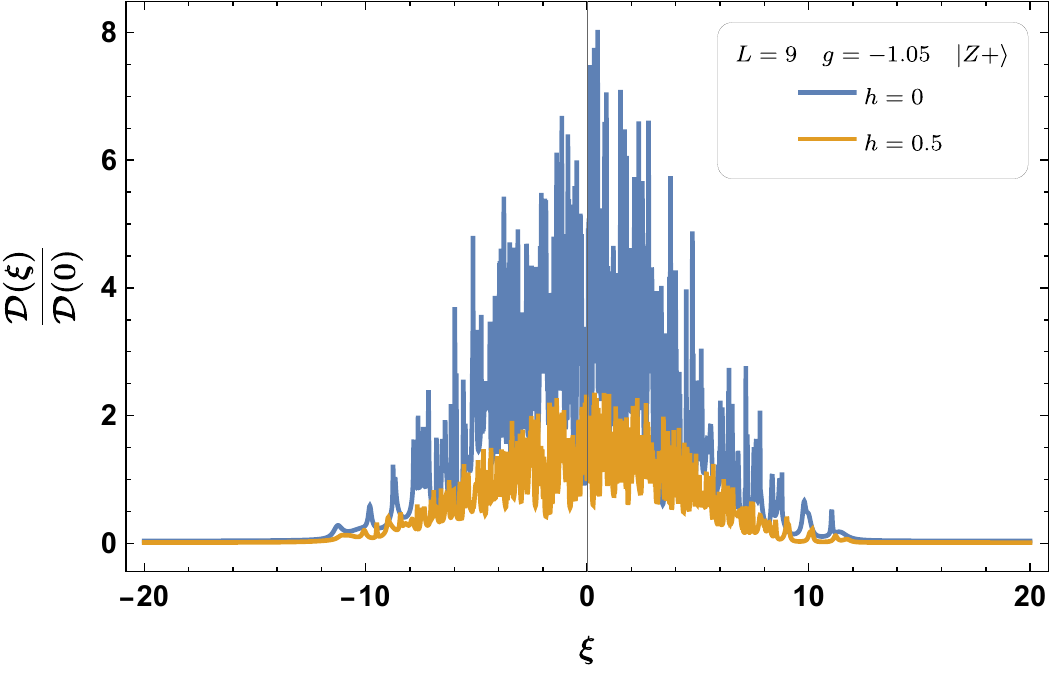}
\caption{
Krylov distribution $\mathcal{D}(\xi)$ for the Ising chain with $L=9$ and
$g=-1.05$ for different product initial states: (left) $|X+\rangle$, (center)
$|Y+\rangle$, and (right) $|Z+\rangle$. Blue curves correspond to the integrable
case ($h=0$), while brown curves correspond to the chaotic case ($h=0.5$).
}
\label{fig:XYZ}
\end{figure}

Figure~\ref{fig:XYZ} shows the Krylov distribution $\mathcal{D}(\xi)$ for three
representative product states corresponding to spin polarization along the $X$,
$Y$, and $Z$ directions. Blue curves correspond to the integrable case ($h=0$),
while brown curves correspond to the chaotic case ($h=0.5$). In all cases,
$\mathcal{D}(\xi)$ rapidly saturates to a constant value when $\xi$ lies far
outside the spectral support of the Hamiltonian, reflecting the exponential
localization of resolvent amplitudes along the Krylov chain when the spectral
parameter is separated from the spectrum by a finite gap.

When $\xi$ lies within or near the many-body spectrum, $\mathcal{D}(\xi)$ exhibits
pronounced oscillations as a function of $\xi$. These oscillations arise from
coherent interference between contributions of different energy eigenstates,
weighted by inverse energy factors $(\xi-E_j)^{-1}$. Although the detailed
oscillatory structure depends on the choice of initial state, this qualitative
distinction persists across all product states considered.

\begin{figure}[t]
\centering
\includegraphics[width=0.46\textwidth]{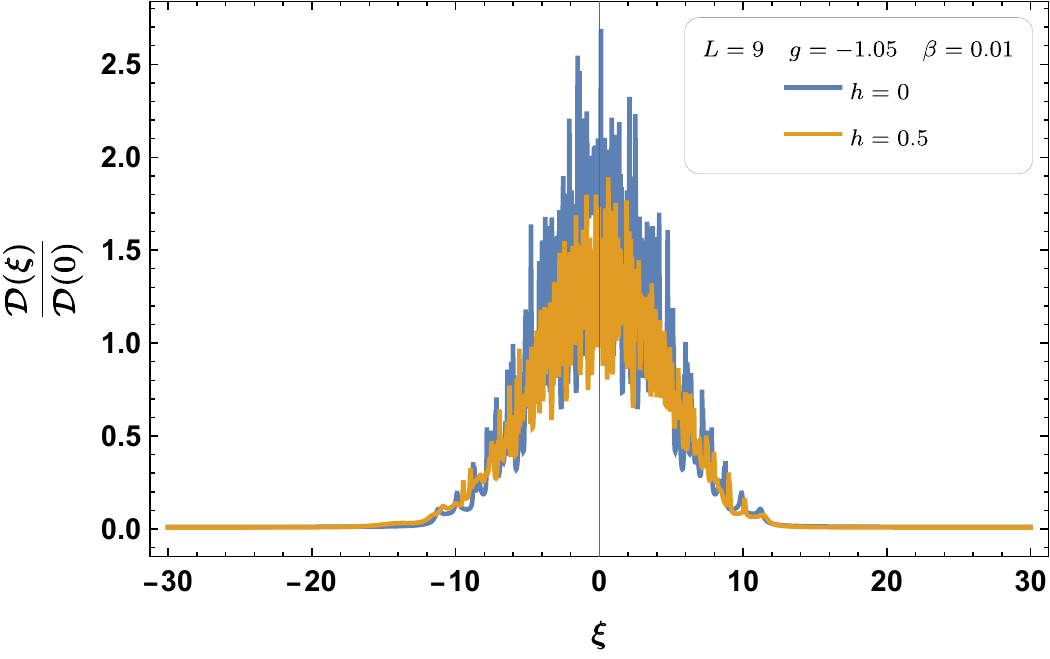}
\caption{
Krylov distribution $\mathcal{D}(\xi)$ for the Gibbs-like initial state
\eqref{eq:TS} with $\beta=0.01$ in an Ising chain of size $L=9$. Blue curve:
integrable case ($h=0$). Brown curve: chaotic case ($h=0.5$).
}
\label{fig:Gibbs}
\end{figure}

Figure~\ref{fig:Gibbs} displays $\mathcal{D}(\xi)$ for the Gibbs-like initial state
\eqref{eq:TS} with $\beta=0.01$. Compared to product states, the distribution is
significantly smoother, reflecting partial self-averaging due to the
superposition over many energy eigenstates spanning a broad portion of the
spectrum.

Beyond spectral sensitivity, our numerical results indicate that the Krylov distribution can encode qualitative distinctions between integrable and chaotic dynamics. Two features are particularly relevant: the amplitude of oscillations in $\mathcal{D}(\xi)$ across the spectrum, and the overall magnitude of $\mathcal{D}(\xi)$ within the spectral region. For product-state initializations, the integrable regime consistently exhibits larger oscillation amplitudes and higher $\mathcal{D}(\xi)$ values, whereas the chaotic regime shows smoother, more suppressed behavior. Even for Gibbs-like initial states, which partially average over the spectrum, the distinction between regimes remains visible, with integrable dynamics producing more pronounced modulations.

\begin{figure}[t]
 \centering 
 \includegraphics[width=0.46\textwidth]{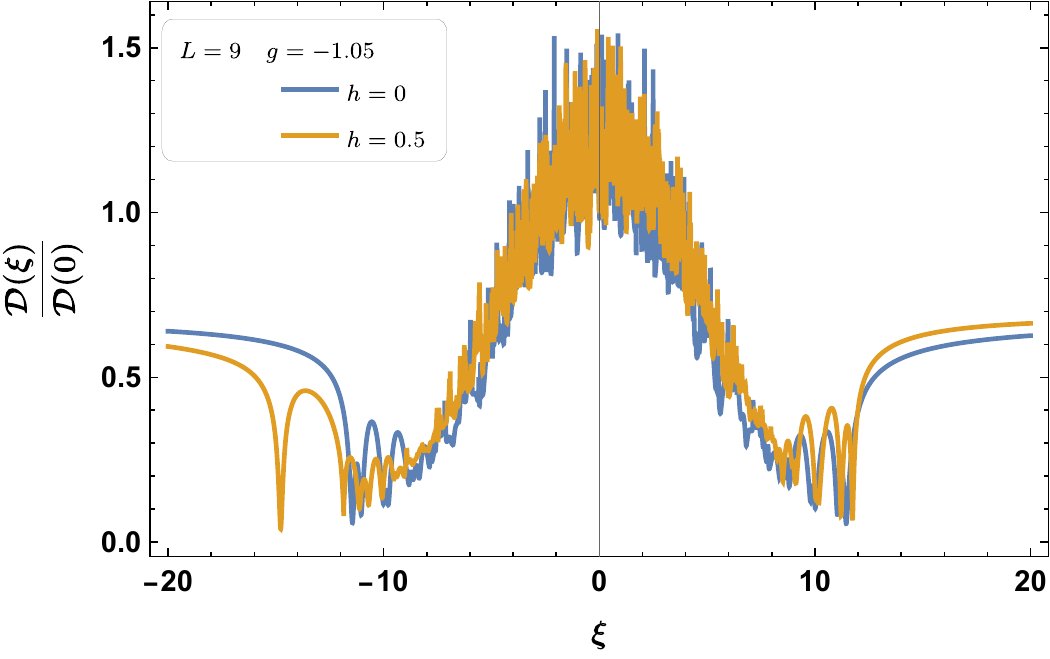}
  \caption{Krylov distribution $\mathcal{D}(\xi)$ averaged over twenty random initial states with positive parity in the Ising chain ($L=9, g=-1.05$). Blue curve: integrable case ($h=0$). Brown curve: chaotic case ($h=0.5$). Ensemble averaging reduces the contrast between regimes, though a mild distinction in smoothness and magnitude persists.}
   \label{fig:random}
    \end{figure}

These observations should be interpreted cautiously due to the modest system size ($L=9$) and sensitivity to initial-state choice. To further probe robustness, we computed $\mathcal{D}(\xi)$ averaged over an ensemble of twenty random initial states with fixed positive parity (Figure~\ref{fig:random}). Ensemble averaging substantially reduces the contrast between regimes: while chaotic averages are slightly smoother and lower in magnitude, the sharp distinctions seen for structured initial states largely diminish. This demonstrates that, although $\mathcal{D}(\xi)$ is sensitive to spectral correlations linked to integrability and chaos, its diagnostic power depends on the degree of spectral averaging in the initial state. For broadly sampling states---such as random superpositions or high-temperature Gibbs states---fine-grained spectral differences are partially washed out, yielding a more universal response.

 While our study highlights the primary sensitivity of the Krylov distribution to spectral structure, more extensive numerical work is needed to establish its full utility as a diagnostic for quantum chaos and thermalization.


\section{Conclusions and Outlook}
\label{sec:conclusion}

In this work, we introduced the Krylov distribution $\mathcal{D}(\xi)$ as a static, resolvent-based diagnostic that organizes inverse-energy response in terms of Krylov-space structure. By elevating the resolvent-dressed state $(H-\xi)^{-1}|\psi_0\rangle$ to a primary object, this framework provides a transparent connection between spectral properties, static response functions, and the emergent geometry of Krylov space. Combining asymptotic analysis, exact solutions in analytically tractable models, and numerical studies of interacting systems, we showed that $\mathcal{D}(\xi)$ exhibits robust and interpretable scaling behavior across gapped, gapless, and critical regimes.

A central result is that the Krylov distribution sharply distinguishes whether the spectral parameter $\xi$ lies inside or outside the many-body spectrum. When $\xi$ is separated from the spectrum by a finite gap, the resolvent-dressed state is exponentially localized in Krylov space and $\mathcal{D}(\xi)$ saturates to an $\mathcal{O}(1)$ value in the thermodynamic limit. By contrast, when $\xi$ lies within a continuous part of the spectrum associated with an absolutely continuous spectral measure, resolvent amplitudes remain extended along the Krylov chain in an averaged sense, leading to extensive growth $\mathcal{D}(\xi)\sim N/2$ for an effective Krylov dimension $N$. This spectral dichotomy provides a direct diagnostic that complements conventional probes such as spectral functions and densities of states.

The exactly solvable models illustrate these principles from complementary perspectives. Systems with constant Lanczos coefficients display a smooth interpolation from exponential saturation outside the spectral band to linear growth inside it, capturing the essential features of bounded continuous spectra. The displaced harmonic oscillator, characterized by an unbounded discrete spectrum and square-root-growing coefficients, yields closed-form amplitudes $\psi_n(\xi)$ and a Krylov distribution with sharp resonant peaks at eigenvalues separated by strongly localized regions. The $\mathrm{SU}(1,1)$ chain with linearly growing coefficients provides a tractable model of continuous spectra with asymptotically linear Lanczos growth; in this case, $\mathcal{D}_N(\xi)\sim N/\ln N$ signals power-law delocalization and the absence of spectral gaps. Together, these examples span bounded and unbounded, discrete and continuous spectra and their distinct manifestations in Krylov space.

Our numerical study of the mixed-field Ising chain further shows that while $\mathcal{D}(\xi)$ is primarily sensitive to spectral boundaries, it can also retain signatures of integrability versus chaos for structured initial states. Product states exhibit larger oscillations and enhanced magnitude of $\mathcal{D}(\xi)$ in the integrable regime, whereas chaotic dynamics produce smoother and more suppressed profiles. These distinctions are partially preserved for Gibbs-like superpositions but are substantially reduced under ensemble averaging with random states. Thus, $\mathcal{D}(\xi)$ should not be viewed as a universal chaos indicator, but rather as a tunable probe whose diagnostic power depends on the reference state and the degree of spectral averaging.

Beyond its role as a spectral diagnostic, the Krylov distribution is closely tied to static response and quantum geometry. The fidelity susceptibility corresponds directly to the total weight $\sum_n |\psi_n(\xi)|^2$ of the resolvent-dressed state evaluated at the ground-state energy. More generally, the quantum geometric tensor admits a natural decomposition in terms of Krylov-resolved amplitudes, highlighting the Krylov basis as an efficient representation of quantum geometry and adiabatic response. Higher inverse-gap moments can likewise be generated through derivatives of $\psi_n(\xi)$ with respect to $\xi$, providing a systematic framework for static response functions of increasing order.

At quantum critical points, the Krylov distribution is expected to exhibit characteristic scaling. In gapless systems with finite-size gap $\Delta\sim 1/L$, the exponential energy resolution inherent to Krylov space implies that resolving energies down to $\Delta$ requires a Krylov depth growing logarithmically with system size, naturally leading to $\mathcal{D}(\xi)\sim \log L$ at criticality. A more detailed investigation of this connection may provide a practical route to diagnosing critical behavior in numerical and experimental settings.

Our results also clarify the relationship between static and dynamical Krylov diagnostics. While Krylov complexity $\mathcal{C}(t)$ tracks the spreading of $e^{-iHt}|\psi_0\rangle$ and is sensitive to operator growth and scrambling, the Krylov distribution $\mathcal{D}(\xi)$ encodes how resolvent-dressed states explore Krylov space as a function of energy. These perspectives are related through the Laplace transform,
\begin{equation}
\psi_n(\xi) = -i \int_0^\infty dt\, e^{i\xi t}\, \phi_n(t),
\end{equation}
which connects time-domain spreading to energy-domain resolution. Nevertheless, $\mathcal{D}(\xi)$ probes inverse-energy structure in a manner qualitatively distinct from finite-time dynamical diagnostics, particularly near spectral gaps, edges, and critical points.

More broadly, given a reference state $|\psi_0\rangle$, the Krylov basis generated by repeated applications of $H$ provides a unified framework for analyzing general operator functions $f(H)$. Time evolution and the resolvent correspond to two specific choices, $f_t(H)=e^{-iHt}$ and $f_\xi(H)=(H-\xi)^{-1}$, but many others arise in applications, including transfer functions, exponential and trigonometric propagators, fractional powers, and the sign function (see, for example, \cite{Guttel2017}). In general, any dressed state of the form $f(H)|\psi_0\rangle$ admits a Krylov expansion governed by the same Lanczos coefficients (see, e.g., \cite{Francois:2024rdm,Francois:2025lqn,Francois:2025shu}). The choice of $f$ selects which spectral features are emphasized, while the Krylov geometry organizes how that information is distributed.

Several directions for future work naturally emerge. Quantitative measures derived from fluctuations or correlations of $\mathcal{D}(\xi)$ along the spectral axis may provide additional probes of integrability and chaos beyond traditional level statistics, and systematic finite-size scaling would help assess their universality. Extending the framework to many-body localized systems, where Krylov-space localization may persist even within the spectrum, is another promising direction. Generalizations to
non-Hermitian Hamiltonians or Lindbladian dynamics could shed light on open quantum systems. On the mathematical side, a more rigorous understanding of universal scaling laws—such as $N^{2/3}$ behavior near generic spectral edges—would strengthen connections to random matrix theory and orthogonal polynomial asymptotics.

Finally, although direct experimental access to the full Krylov distribution remains challenging, its connection to fidelity susceptibility and quantum geometric tensors suggests that certain aspects of $\mathcal{D}(\xi)$ may be accessible in quantum simulators using existing measurement protocols.

In summary, we have established the Krylov distribution $\mathcal{D}(\xi)$ as a static counterpart to dynamical Krylov complexity. By linking spectral structure, scaling behavior, quantum geometry, and many-body dynamics within a unified Krylov-space framework, this work lays the foundation for systematically probing energy-resolved structure in quantum many-body systems.

\section*{Acknowledgments}

We would like to thank Souvik Banerjee and  Mohammad Reza Tanhayi for 
many insightful discussions on various aspects of  
Krylov space and Krylov complexity. The work of 
M. A.  is supported by the Iran National Science Foundation (INSF) under 
Project No.~4023620.  We also
acknowledge the assistance of ChatGPT for help to improve the quality of the written text.


\appendix

\section{Mathematical Properties of the Krylov Distribution} 
\label{app:bounds}

In this appendix, we analyze the fundamental properties of the Krylov distribution
\begin{equation}
\mathcal{D}(\xi) = \sum_{n=0}^{d_\psi-1} n\, P_n(\xi), 
\qquad
P_n(\xi) = \frac{|\psi_n(\xi)|^2}{\sum_{\ell=0}^{d_\psi-1} |\psi_\ell(\xi)|^2},
\end{equation}
where the Krylov resolvent amplitudes are defined as
\begin{equation}
\psi_n(\xi) = \langle n | (H-\xi)^{-1} | \psi_0 \rangle.
\end{equation}
Since $P_n(\xi)$ is a normalized probability distribution over the Krylov indices $n=0,1,\dots,d_\psi-1$, the Krylov distribution $\mathcal{D}(\xi)$ can be interpreted as the average position along the Krylov chain. By definition, as a weighted average of integers bounded between $0$ and $d_\psi-1$, it satisfies
\begin{equation}
0 \le \mathcal{D}(\xi) \le d_\psi-1.
\end{equation}

However, these extremal values are generically unattainable in finite chains with nonzero off-diagonal Lanczos coefficients. The amplitudes $\psi_n(\xi)$ satisfy the tridiagonal recurrence relation
\begin{equation}
b_{n+1} \psi_{n+1} + (a_n-\xi) \psi_n + b_n \psi_{n-1} = \delta_{n0}, 
\end{equation}
with $\psi_{-1}=\psi_{d_\psi}=0$ and $b_n>0$, which enforces nearest-neighbor coupling along the Krylov chain. As a result, the probability distribution $\{P_n\}$ cannot collapse entirely onto a single site.

To make this constraint quantitative, define the normalized amplitudes
\begin{equation}
\tilde{\psi}_n = \frac{\psi_n}{\sqrt{\sum_\ell |\psi_\ell|^2}},
\qquad
P_n = |\tilde{\psi}_n|^2 ,
\end{equation}
which satisfy the homogeneous recurrence
\begin{equation}
b_{n+1}\tilde{\psi}_{n+1} + b_n\tilde{\psi}_{n-1} = (\xi-a_n)\tilde{\psi}_n .
\label{eq:recurrence_normalized}
\end{equation}
Taking absolute values and applying the triangle inequality gives
\begin{equation}
b_n|\tilde{\psi}_{n-1}| + b_{n+1}|\tilde{\psi}_{n+1}| \ge |\xi-a_n|\,|\tilde{\psi}_n| .
\label{eq:triangle_ineq}
\end{equation}
Defining
\begin{equation}
\kappa_n \equiv \frac{|\xi-a_n|}{b_n+b_{n+1}},
\end{equation}
we obtain the bound
\begin{equation}
\max(P_{n-1},P_{n+1}) \ge \kappa_n^2\, P_n .
\label{eq:prob_bound}
\end{equation}

This equation immediately yields nontrivial bounds on the Krylov distribution itself. Because perfect single-site localization is forbidden, the most left-skewed admissible distribution must involve weight on sites $n=0$ and $n=1$. Applying Eq.~(\ref{eq:prob_bound}) at $n=0$ gives
\begin{equation}
P_1 \ge \frac{|\xi-a_0|^2}{b_1^2}\, P_0 .
\end{equation}
Using normalization, this implies
\begin{equation}
P_1 \ge \frac{\kappa_0^2}{1+\kappa_0^2},
\qquad
\kappa_0 \equiv \frac{|\xi-a_0|}{b_1}.
\end{equation}
Since $\mathcal{D}(\xi)=\sum_n nP_n \ge P_1$, we obtain the sharp lower bound
\begin{equation}
\mathcal{D}(\xi) \ge \frac{\kappa_0^2}{1+\kappa_0^2}.
\label{eq:sharp_lower_bound}
\end{equation}

A completely analogous argument applies at the opposite end of the Krylov chain. The most right-skewed admissible distribution must involve sites $n=d_\psi-2$ and $n=d_\psi-1$. Applying Eq.~(\ref{eq:prob_bound}) at $n=d_\psi-1$ yields
\begin{equation}
P_{d_\psi-2} \ge \frac{|\xi-a_{d_\psi-1}|^2}{b_{d_\psi-1}^2}\, P_{d_\psi-1}.
\end{equation}
Defining
\begin{equation}
\kappa_{d_\psi-1} \equiv \frac{|\xi-a_{d_\psi-1}|}{b_{d_\psi-1}},
\end{equation}
and using $\mathcal{D}(\xi)\le (d_\psi-1)-P_{d_\psi-2}$, we obtain the sharp upper bound
\begin{equation}
\mathcal{D}(\xi)
\le (d_\psi-1)
- \frac{\kappa_{d_\psi-1}^2}{1+\kappa_{d_\psi-1}^2}.
\label{eq:sharp_upper_bound}
\end{equation}

Together, Eqs.~(\ref{eq:sharp_lower_bound}) and~(\ref{eq:sharp_upper_bound}) show that the Krylov distribution is confined to a strict interior interval of the Krylov chain. The width of the forbidden boundary layers is controlled by the local detuning of the spectral parameter $\xi$ from the diagonal elements and by the corresponding Lanczos coefficients. This provides a precise quantitative expression of the intuitive statement that resolvent-dressed states must exhibit a minimum degree of spreading in Krylov space.

A particularly interesting scenario occurs if the probabilities are approximately uniform across the chain. In the hypothetical limit where all $P_n$ are exactly equal, $P_n = 1/d_\psi$, the Krylov distribution attains the midpoint
\begin{equation}
\mathcal{D}(\xi) = \sum_{n=0}^{d_\psi-1} n\, \frac{1}{d_\psi} = \frac{d_\psi-1}{2}.
\end{equation}

Exact uniformity is impossible in finite chains with generic $b_n>0$ due to the inhomogeneous term at $n=0$ and the boundary at $n=d_\psi-1$, which break translational symmetry and typically induce exponential decay/growth of amplitudes from the boundaries. Nevertheless, in sufficiently large chains and for spectral parameters $\xi$ far from resonances, the amplitudes $|\psi_n(\xi)|^2$ can become nearly uniform over a significant portion of the chain. In this quasi-uniform regime, the Krylov distribution approaches the midpoint $(d_\psi-1)/2$, reflecting a state that is spread roughly evenly along the Krylov chain. This demonstrates that while the extremes are forbidden, the central region of the chain remains physically accessible and can dominate the average. 

Physically, these constraints highlight a fundamental property of the resolvent operator $(H-\xi)^{-1}$. Even if the initial state coincides with a single Krylov basis vector, the action of the resolvent inevitably mixes adjacent Krylov layers due to the off-diagonal elements $b_n$, and the resulting distribution cannot collapse entirely to a single site. The Krylov distribution therefore always reflects a minimum degree of spreading along the chain, with the precise location determined by the interplay of the spectral parameter $\xi$ and the Lanczos coefficients. Values of $\mathcal{D}(\xi)$ near the midpoint correspond to quasi-uniform distributions and indicate significant delocalization along the chain, while values closer to the lower or upper ends signify a skewing toward the beginning or end of the chain, respectively.

\section{Krylov Resolution of Quantum Geometry}
\label{app:fidelity}

This appendix develops the relationship between Krylov space and quantum geometry, providing the mathematical foundation for the connection discussed in the main text. We begin with the standard formulation in the energy eigenbasis and then generalize to arbitrary initial states, showing how geometric response functions admit natural decompositions across Krylov layers.

For a parameter-dependent Hamiltonian $H(\vec{\lambda})$ with a nondegenerate ground state $|E_0(\vec{\lambda})\rangle$, the standard quantum geometric tensor is defined in the energy eigenbasis as
\begin{equation}
Q_{\mu\nu}^{\text{std}}(\vec{\lambda}) = \sum_{n \neq 0} \frac{\langle E_0|\partial_\mu H|E_n\rangle \langle E_n|\partial_\nu H|E_0\rangle}{(E_n - E_0)^2},
\end{equation}
which reduces to the fidelity susceptibility $\chi_F^{\mu\nu}$ for diagonal components. This formulation explicitly references the energy eigenbasis $\{|E_n\rangle\}$ and depends crucially on the choice of the ground state.

To extend this framework to arbitrary initial states $|\psi_0\rangle$ and to make explicit its organization relative to dynamical accessibility, we proceed as follows. For a given initial state $|\psi_0\rangle$ (which need not be an energy eigenstate) and a spectral parameter $\xi \in \mathbb{C}$ chosen such that the resolvent $(H-\xi)^{-1}$ is well defined, we generate the Krylov basis $\{|n\rangle\}$ from $|\psi_0\rangle$ via the Lanczos algorithm. This basis spans the Krylov subspace
\[
\mathcal{K} = \text{span}\{|\psi_0\rangle, H|\psi_0\rangle, H^2|\psi_0\rangle, \dots\},
\]
representing the part of Hilbert space dynamically accessible from $|\psi_0\rangle$ under unitary evolution. We then define the Krylov resolvent amplitudes
\begin{equation}
\psi_n^{(\mu)}(\vec{\lambda}, \xi) = \langle n | (H(\vec{\lambda}) - \xi)^{-1} \partial_\mu H(\vec{\lambda}) | \psi_0 \rangle,
\end{equation}
where $\partial_\mu \equiv \partial/\partial\lambda^\mu$. These amplitudes measure how parameter perturbations, filtered through the resolvent at spectral parameter $\xi$, project onto different Krylov layers. From these amplitudes we construct the $\xi$-dependent Krylov quantum geometric tensor
\begin{equation}
Q_{\mu\nu}^{\psi_0}(\vec{\lambda}, \xi) = \sum_{n \ge 0} [\psi_n^{(\mu)}(\vec{\lambda}, \xi)]^* \psi_n^{(\nu)}(\vec{\lambda}, \xi).
\end{equation}

When the Krylov subspace $\mathcal{K}$ is sufficiently large—specifically, when it contains the support of $(H-\xi)^{-1}\partial_\mu H|\psi_0\rangle$ for all $\mu$—and upon inserting the completeness relation $I_{\mathcal{K}}=\sum_n |n\rangle\langle n|$ over $\mathcal{K}$, this expression can be written as
\begin{equation}
Q_{\mu\nu}^{\psi_0}(\vec{\lambda}, \xi)
=
\langle \psi_0 | \partial_\mu H^\dagger (H-\xi^*)^{-1} (H-\xi)^{-1} \partial_\nu H | \psi_0 \rangle.
\end{equation}
This provides a resolvent representation of the geometric response function. For $\xi=E_0$ and $|\psi_0\rangle=|E_0\rangle$, and with the understanding that the ground-state pole is removed via the reduced resolvent, this expression reduces to the standard ground-state fidelity susceptibility.

A subtlety arises when $|\psi_0\rangle$ is exactly an energy eigenstate. In this case, repeated application of $H$ generates a trivial Krylov space consisting only of $|\psi_0\rangle$ itself. Moreover, the reduced resolvent $R_0=(H-E_0)^{-1}P$, with $P$ projecting out the ground state, produces states orthogonal to $|E_0\rangle$, which therefore have vanishing expansion coefficients in the one-dimensional Krylov basis $\{|E_0\rangle\}$. This reflects the collapse of the Krylov subspace rather than a failure of the formalism, and highlights that the Krylov decomposition is most naturally suited to initial states that are not exact energy eigenstates. Such states are physically relevant in experimental settings, where one typically prepares simple product states, coherent states, or other non-eigenstates.

The physical interpretation of the Krylov decomposition is twofold. First, the spectral parameter $\xi$ controls energy resolution: choices of $\xi$ near the ground-state energy probe low-energy response (in the reduced-resolvent sense), values of $\xi$ in spectral gaps probe virtual excitations, and values within continuous spectra characterize extended or scattering-like response. Second, the sum over Krylov index $n$ reveals how the geometric response distributes across Krylov layers, providing a static analogue of the Krylov distribution $\mathcal{D}(\xi)$ introduced in the main text. The weights $|\psi_n^{(\mu)}(\xi)|^2$ quantify how sensitivity to parameter changes is organized by dynamical accessibility.

Further insight is obtained from the recurrence relations satisfied by the amplitudes. Projecting the identity $(H-\xi)(H-\xi)^{-1}=I$ onto the Krylov basis yields
\begin{equation}
b_{n+1} \psi_{n+1}^{(\mu)} + (a_n - \xi) \psi_n^{(\mu)} + b_n \psi_{n-1}^{(\mu)} = \mathcal{O}_{\mu,n},
\end{equation}
where $\mathcal{O}_{\mu,n}=\langle n | \partial_\mu H | \psi_0 \rangle$ are the bare matrix elements of the perturbation in the Krylov basis. This equation shows how the resolvent dresses the bare perturbation as it propagates along the Krylov chain, with the source terms $\mathcal{O}_{\mu,n}$ encoding the direct coupling of the perturbation to different Krylov layers.

For the specific case of ground-state fidelity susceptibility, alternative Krylov constructions are possible. One approach generates the Krylov basis not from $|E_0\rangle$ itself but from the perturbed state $|\phi\rangle=\partial_\lambda H|E_0\rangle$, provided it is not proportional to $|E_0\rangle$. For a single parameter $\lambda$, this yields a valid decomposition $\chi_F=\sum_n |\psi_n^{\partial H}|^2$, where $\psi_n^{\partial H}=\langle n|R_0 \partial_\lambda H|E_0\rangle$ and the Krylov basis $\{|n\rangle\}$ is generated from $|\phi\rangle$. For multiple parameters, however, this construction leads to different Krylov bases for different $\mu$, so that off-diagonal components $Q_{\mu\nu}$ involve overlaps between distinct bases. One may instead generate a common Krylov basis from a suitable superposition $\sum_\mu c_\mu \partial_\mu H|E_0\rangle$. Finally, it is worth noting that the energy eigenbasis itself can be viewed as a trivial Krylov decomposition, with each eigenstate forming a one-dimensional Krylov subspace, in which case the above expressions reproduce the standard quantum geometric tensor exactly.

Overall, this framework shows that quantum geometry admits a natural organization relative to dynamical accessibility from a chosen initial state. Decomposition across Krylov layers reveals how different regions of the dynamically accessible subspace contribute to parameter sensitivity. This perspective is particularly well suited for analyzing the response of experimentally relevant, non-eigenstate preparations, and highlights the complementary roles played by spectral resolution through $\xi$ and Krylov-space structure through the index $n$ in shaping quantum geometric response.


\section{Resolvent Krylov Basis and Generalized Krylov Constructions} 
\label{app:res_krylov}

In the main text, we analyzed resolvent-dressed states using the standard Krylov basis generated by powers of the Hamiltonian $H$. However, alternative Krylov constructions adapted specifically to inverse-energy physics can provide complementary insights. In this appendix, we develop the theory of the resolvent Krylov basis, generated by powers of $H^{-1}$, and discuss its properties and potential applications.

The standard Krylov basis $\{|n\rangle\}$ is generated by the sequence $\{H^n|\psi_0\rangle\}_{n=0}^\infty$, emphasizing high-energy features through large powers of $H$. For studying low-energy physics or inverse-gap phenomena, it may be advantageous to use a basis that naturally weights low energies more heavily. This motivates considering the sequence $\{H^{-n}|\psi_0\rangle\}_{n=1}^\infty$, which corresponds to repeated applications of the inverse Hamiltonian.

A motivation to study this space is as follows. Assuming that the Hamiltonian $H$ is invertible and that
the reference state $|\psi_0\rangle$ has nonvanishing overlap with low-energy
eigenstates, we may formally expand the resolvent-dressed state as
\begin{equation}
    |\psi(\xi)\rangle
    = (H - \xi)^{-1} |\psi_0\rangle
    = \sum_{n=0}^{\infty} \xi^{\,n} H^{-n-1} |\psi_0\rangle,
\end{equation}
which is valid for $|\xi| \ll \|H\|$. This expansion motivates the definition of
the resolvent Krylov space
\begin{equation}\label{eq:res_Krylov_space}
    \mathcal{K}_{\mathrm{res}}
    = \mathrm{span}\big\{  H^{-1}|\psi_0\rangle,\,  H^{-2}|\psi_0\rangle,\,
        H^{-3}|\psi_0\rangle,\,   \dots
      \big\}.
\end{equation}
Note that $|\psi_0\rangle$ itself is not included in this span unless it lies in the range of $H^{-1}$, which is typically not the case for generic $|\psi_0\rangle$. The dimension $d_{\mathrm{res}} \leq \mathcal{D}-1$ may differ from the dimension $d_\psi$ of the standard Krylov space.

Unlike the standard Krylov space generated by powers of $H$, the reference state
$|\psi_0\rangle$ itself does not appear as a basis element in
$\mathcal{K}_{\mathrm{res}}$. This reflects the intrinsically static nature of
resolvent physics, which probes inverse energy scales rather than dynamical
trajectories.

To construct an orthonormal basis for $\mathcal{K}_{\mathrm{res}}$, we apply the
Lanczos algorithm to the operator $H^{-1}$. The first basis vector is defined as
\begin{equation}
    |r_1\rangle
    = \frac{H^{-1}|\psi_0\rangle}
           {\|H^{-1}|\psi_0\rangle\|}.
\end{equation}
Setting $|0\rangle \equiv 0$ and $\beta_1 \equiv 0$, the remaining basis vectors
are generated recursively for $n \ge 1$ via
\begin{equation}
    \widehat{|r_{n+1}\rangle}
    = (H^{-1} - \alpha_n)|r_n\rangle
      - \beta_n |r_{n-1}\rangle,
    \qquad
    |r_{n+1}\rangle
    = \frac{\widehat{|r_{n+1}\rangle}}{\beta_{n+1}},
    \label{eq:res_lanczos}
\end{equation}
with Lanczos coefficients
\begin{equation}
    \alpha_n = \langle r_n | H^{-1} | r_n \rangle,
    \qquad
    \beta_{n+1} = \|\widehat{|r_{n+1}\rangle}\|.
\end{equation}
The procedure terminates when $\beta_{n+1}=0$, assumed to occur at $n=d_\psi$,
which defines the dimension of the resolvent Krylov subspace. By construction,
the basis is orthonormal,
\begin{equation}
    \langle r_m | r_n \rangle = \delta_{mn},
\end{equation}
and the operator $H^{-1}$ is tridiagonal in this basis.

The resolvent-dressed state can now be expanded as
\begin{equation}
    |\psi(\xi)\rangle
    = \sum_{n=1}^{d_\psi} \chi_n(\xi)\, |r_n\rangle,
\end{equation}
where the resolvent Krylov amplitudes are
\begin{equation}
    \chi_n(\xi)
    \equiv \langle r_n | (H - \xi)^{-1} | \psi_0 \rangle.
\end{equation}
Projecting the resolvent equation $(H-\xi)|\psi(\xi)\rangle = |\psi_0\rangle$
onto the resolvent Krylov basis yields the exact recursion relation (for
$\xi \neq 0$)
\begin{align}
 (\xi^{-1} - \alpha_n)\,\chi_n(\xi)
    &- \beta_{n+1}\,\chi_{n+1}(\xi)
    - \beta_n\,\chi_{n-1}(\xi)
    \notag\\ &=
    \xi^{-1} \langle r_n | H^{-1} | \psi_0 \rangle,
    \qquad n \ge 1,
\end{align}
with $\chi_0(\xi)\equiv 0$. For $n \ge 2$, the inhomogeneous term vanishes and the
recursion becomes homogeneous.

Finally, inserting a complete set of energy eigenstates
$H|E_\ell\rangle = E_\ell |E_\ell\rangle$ yields the spectral representation
\begin{equation}
    \chi_n(\xi)
    = \sum_\ell
      \frac{\langle r_n | E_\ell\rangle
            \langle E_\ell | \psi_0\rangle}
           {E_\ell - \xi}.
\end{equation}
From this expression it is clear that low-energy states with
$|E_\ell - \xi| \ll 1$ are strongly enhanced, while high-energy contributions are
suppressed. Consequently, the resolvent Krylov basis provides a complementary
resolution of spectral weight compared to the standard Krylov basis, effectively
probing the spectrum through the inverse-energy map $E \mapsto E^{-1}$.
Note that in this notation the Krylov distribution is given by
\begin{equation}
\mathcal{D}(\xi)=\sum_{n\ge0}\,n|\chi_n(\xi)|^2\,.
\end{equation}

More generally, one may consider extended Krylov constructions
generated by the joint action of $H$ and $H^{-1}$ on $|\psi_0\rangle$ \cite{Druskin:1998,Jagels:2009,Daas:2025}, thereby interpolating between dynamical and resolvent-based descriptions. A systematic
analysis of such hybrid constructions is left for future work.


\section{Spectral Edge and Critical Points: Details}
\label{app:asymptotics}

Consider a spectral edge at $E_*$ where the density of states vanishes as a power law,
\begin{equation}
    \rho_0(E) \sim C \, (E - E_*)^\alpha, \quad E \gtrsim E_*, \quad \alpha > -1.
    \label{eq:rho_power}
\end{equation}
For generic band edges in systems with quadratic dispersion, $\alpha = 1/2$, while other exponents may arise from non-generic edges or from different asymptotic behaviors of the underlying Jacobi coefficients. We distinguish two regimes: (i) regular (turning-point) edges where the group velocity vanishes, and (ii) non-turning-point edges where the group velocity remains finite.

At a regular spectral edge where the recurrence coefficients approach constants, $a_n\to a_\infty$ and $b_n\to b_\infty$, the classical frequency
\begin{equation}
\omega(E) = 2b_\infty \sqrt{1 - \left(\frac{E-a_\infty}{2b_\infty}\right)^2}
\end{equation}
vanishes as $\omega(E) \sim \sqrt{E-E_*}$. This vanishing group velocity creates a turning point that governs the asymptotic behavior of the orthogonal polynomials.

To derive the universal scaling, we approximate $Q_n(E)\equiv Q(n)$ as a smooth function of $n$ and expand the Jacobi recurrence
\begin{equation}
b_{n+1}Q_{n+1}(E) + a_n Q_n(E) + b_n Q_{n-1}(E) = E Q_n(E)
\end{equation}
for large $n$ near $E = E_* + \delta E$. Using $Q_{n\pm1}=Q(n)\pm Q'(n)+\tfrac12 Q''(n)+\cdots$ and $a_n\simeq a_\infty$, $b_n\simeq b_\infty$, the linear derivative terms cancel, yielding to leading order
\begin{equation}
b_\infty Q''(n) + \delta E \, Q(n) \approx 0.
\label{eq:continuum_edge}
\end{equation}

Assuming $\delta E\sim n^{-v}$ and that $Q(n)$ varies on a scale $n^\delta$ (so that $Q''\sim n^{-2\delta}Q$), balancing terms in (\ref{eq:continuum_edge}) gives $2\delta=v$. Requiring self-consistency of the continuum approximation fixes $\delta=1/3$, hence $v=2/3$. Introducing the scaling variable
\begin{equation}
x = c \, n^{2/3}(E-E_*), \qquad c=b_\infty^{-1/3},
\end{equation}
and the ansatz $Q_n(E)=n^{-1/3}\phi(x)$, Eq.~(\ref{eq:continuum_edge}) reduces in the $n\to\infty$ limit to the Airy equation
\begin{equation}
\phi''(x)-x\,\phi(x)=0,
\end{equation}
leading to the universal scaling form
\begin{equation}
Q_n\!\left(E_*+\frac{x}{c n^{2/3}}\right)\sim A\,n^{-1/3}\,\mathrm{Ai}(x).
\label{eq:Q_airy_universal}
\end{equation}

To probe the resolvent at the edge, one must consider the corresponding edge-scaling regime. Setting
\begin{equation}
\xi = E_* + \frac{i\eta}{c n^{2/3}}, \qquad \eta>0,
\end{equation}
ensures that the imaginary part competes with the level spacing in the critical region. Substituting (\ref{eq:Q_airy_universal}) and (\ref{eq:rho_power}) with $\alpha=1/2$ into the spectral representation
\begin{equation}
\psi_n(\xi)=\int_{E_*}^\infty \frac{Q_n(E)}{\xi-E}\,\rho_0(E)\,dE,
\end{equation}
and changing variables to $x=c n^{2/3}(E-E_*)$, we find
\begin{equation}
\psi_n(\xi)\sim n^{-2/3}\int_0^\infty \frac{\mathrm{Ai}(x)\sqrt{x}}{i\eta-x}\,dx.
\end{equation}
The integral converges, yielding $|\psi_n(\xi)|^2\sim n^{-4/3}$. The edge contribution then gives
\begin{equation}
\sum_{n=1}^N |\psi_n(\xi)|^2\sim\text{const},\qquad
\sum_{n=1}^N n|\psi_n(\xi)|^2\sim N^{2/3},
\end{equation}
leading to the universal edge scaling
\begin{equation}
\mathcal{D}(\xi)\sim N^{2/3}.
\label{eq:airy_scaling}
\end{equation}
This sublinear growth reflects critical slowing near regular spectral edges, where the group velocity vanishes.

For edges where the group velocity remains finite, $\theta'(E_*)\neq0$, the orthogonal polynomials exhibit oscillatory asymptotics,
\begin{equation}
Q_n(E)\sim\cos\bigl(n\theta(E)+\phi(E)\bigr).
\end{equation}
Linearizing $\theta(E)\approx c(E-E_*)$ and assuming $\xi\neq E_*$, the resolvent amplitude scales as
\begin{equation}
|\psi_n(\xi)|^2\sim n^{-2(1+\alpha)},
\label{eq:non_turning_point_scaling}
\end{equation}
up to oscillatory prefactors.

Writing $|\psi_n|^2\sim n^{-\beta}$ with $\beta=2(1+\alpha)$, the Krylov distribution behaves as
\begin{equation}
\mathcal{D}(\xi)\sim
\frac{\sum_{n=1}^N n^{1-\beta}}{\sum_{n=1}^N n^{-\beta}},
\end{equation}
which yields
\begin{equation}
\mathcal{D}(\xi)\sim
\begin{cases}
\text{constant}, & \alpha>0,\\
\ln N, & \alpha=0,\\
N^{-2\alpha}, & -\tfrac12<\alpha<0,\\
N/\ln N, & \alpha=-\tfrac12,\\
N, & \alpha<-\tfrac12.
\end{cases}
\label{eq:D_scaling_general}
\end{equation}

The turning-point case $\alpha=1/2$ lies outside this oscillatory classification and yields the distinct Airy scaling $\mathcal{D}(\xi)\sim N^{2/3}$. Together, these results establish a direct connection between the edge behavior of the spectral density and the asymptotic growth of the Krylov distribution.


\section{Quadratic Hamiltonian: Exact  Krylov Analysis}
\label{app:details}

In this appendix we present a complete and exact analysis of the Krylov resolvent amplitudes for the quadratic (displaced oscillator) Hamiltonian. All steps are carried out explicitly, with careful attention to analytic continuation, convergence, and sign conventions. This model provides a rare example where the Krylov dynamics can be solved in closed form.

Following \cite{Balasubramanian:2022tpr} we consider the quadratic Hamiltonian
\begin{equation}
H = \omega a^\dagger a + \lambda (a + a^\dagger),
\label{eq:quadratic_hamiltonian}
\end{equation}
where $N = a^\dagger a$ is the number operator. This Hamiltonian can be diagonalized by the displacement operator
\begin{equation}
D(\alpha) = e^{-\gamma a^\dagger +\gamma^* a}, 
\qquad 
\gamma = \frac{\lambda}{\omega}.
\end{equation}
Defining the shifted annihilation operator $\tilde a = a -\gamma$, we obtain
\begin{equation}
H = \omega \tilde a^\dagger \tilde a  - \gamma^2{\omega}.
\end{equation}
The exact spectrum is therefore
\begin{equation}
E_m = \omega m  - \gamma^2\omega, 
\qquad 
m = 0,1,2,\dots\,.
\label{eq:spectrum}
\end{equation}
The eigenstates are displaced Fock states,
\begin{equation}
|E_m\rangle = D(-\gamma)|m\rangle .
\end{equation}

We take the reference state to be the vacuum $|\psi_0\rangle = |0\rangle$. The time-evolved state can be computed exactly using the displacement-operator algebra:
\begin{equation}
|\psi(t)\rangle 
= e^{-iHt}|0\rangle
= e^{i\gamma^2\omega t}
D\!\left(-\gamma \bigl(1-e^{-i\omega t}\bigr)\right)|0\rangle .
\end{equation}
Expanding in the Fock basis (which coincides exactly with the Krylov basis for this Hamiltonian), we obtain the Krylov amplitudes \cite{Balasubramanian:2022tpr}
\begin{equation}
\phi_n(t) = e^{\alpha(t)}\frac{z(t)^n}{\sqrt{n!}},
\qquad 
n=0,1,2,\dots,
\label{eq:phi_n_exact}
\end{equation}
with
\begin{align}
z(t) = -\gamma\bigl(1-e^{-i\omega t}\bigr),\;\;\;\;\alpha(t) =  \gamma^2 \bigl(i\omega t - 1 + e^{-i\omega t}\bigr).
\end{align}
These amplitudes satisfy $\sum_{n=0}^\infty |\phi_n(t)|^2 = 1$ for all $t$. The corresponding Lanczos
 coefficients are \cite{Balasubramanian:2022tpr}
 \begin{equation}
 a_n=\omega n\,,\;\;\;\;\;\;\;b_n=\lambda\sqrt{n}\,.
 \end{equation}

The Krylov resolvent amplitudes may be computed by the Laplace transform
\begin{equation}
\psi_n(\xi) 
= -i \int_0^\infty dt\, e^{i\xi t}\, \phi_n(t),
\qquad 
\Im\xi > 0,
\label{eq:laplace_def}
\end{equation}
which ensures convergence. When evaluating expressions at real $\xi$, all results are understood as limits $\xi \to \xi + i0^+$.

Substituting \eqref{eq:phi_n_exact} gives
\begin{equation}
\psi_n(\xi)
= -i\frac{(-\gamma)^n}{\sqrt{n!}}
\int_0^\infty dt\, e^{i\xi t} e^{\alpha(t)}(1-e^{-i\omega t})^n .
\label{eq:psi_integral}
\end{equation}
Defining the shifted spectral parameter $\tilde\xi \equiv \xi + \gamma^2\omega$, 
we obtain
\begin{equation}
\psi_n(\xi)
= -i\frac{(-\gamma)^n}{\sqrt{n!}e^{\gamma^2}} 
\int_0^\infty dt\, e^{i\tilde\xi t} e^{\gamma^2 e^{-i\omega t}} (1-e^{-i\omega t})^n .
\label{eq:psi_simplified}
\end{equation}
We now make the change of variables
\begin{equation}
u = e^{-i\omega t},
\qquad 
dt = \frac{i}{\omega}\frac{du}{u}.
\end{equation}
The convergence of the original time integral is guaranteed by $\Im\xi>0$, which allows the contour to be deformed so that $u$ runs along the real interval $[0,1]$. The integral becomes
\begin{equation}
\psi_n(\xi)
= \frac{(-\gamma)^n}{\omega\sqrt{n!} e^{\gamma^2}} 
\int_0^1 u^{-\tilde\xi/\omega-1} e^{\gamma^2 u} (1-u)^n du .
\label{eq:psi_integral_u}
\end{equation}
We now parametrize the spectral variable as
\begin{equation}
\xi = \omega\ell  - \gamma^2\omega ,
\end{equation}
where $\ell$ is treated as a complex parameter. Then $\tilde\xi = \omega\ell$, and
\begin{equation}
\psi_n(\ell)
= \frac{(-\gamma)^n}{\omega\sqrt{n!} e^{\gamma^2}} 
\int_0^1 u^{-\ell-1} e^{\gamma^2 u} (1-u)^n du .
\label{eq:psi_beta_integral}
\end{equation}

The integral is initially convergent for $\Re\ell<0$ and can be analytically continued elsewhere. Using the identity
\begin{equation}
\int_0^1 u^{-\ell-1} e^{\gamma^2 u} (1-u)^n du
= B(-\ell,n+1)\,
{}_1F_1(-\ell,n+1-\ell,\gamma^2),
\end{equation}
where  $B(x,y)=\Gamma(x)\Gamma(y)/\Gamma(x+y)$  is the Beta function and ${}_1F_1$ is the  Confluent hypergeometric function. Thus, we obtain
\begin{equation}
\psi_n(\ell)
= \frac{(-\gamma)^n}{\omega\sqrt{n!}e^{\gamma^2}} 
\frac{\Gamma(-\ell)\Gamma(n+1)}{\Gamma(n+1-\ell)}
{}_1F_1(-\ell,n+1-\ell,\gamma^2).
\end{equation}
Using $\Gamma(n+1)=n!$, this simplifies to
\begin{equation}
\psi_n(\ell)
= \frac{(-\gamma)^n }{\omega e^{\gamma^2}}
\frac{\Gamma(-\ell)\sqrt{n!}}{\Gamma(n+1-\ell)}
{}_1F_1(-\ell,n+1-\ell,\gamma^2).
\label{eq:psi_exact_final}
\end{equation}

For $\ell=\epsilon\to0^+$, we use ${}_1F_1(0,n+1,\gamma^2)=1$ and obtain
\begin{equation}
\psi_n(\epsilon)
\approx \frac{(-\gamma)^n}{\omega e^{\gamma^2}\sqrt{n!}}\frac{1}{\epsilon},
\end{equation}
leading to
\begin{equation}
P_n(0)
= \lim_{\epsilon\to0^+}
\frac{|\psi_n(\epsilon)|^2}{\sum_k|\psi_k(\epsilon)|^2}
= e^{-\gamma^2}\frac{\gamma^{2n}}{n!}.
\end{equation}
The Krylov distribution at resonance is therefore
\begin{equation}
\mathcal D(0) = \sum_n nP_n = \gamma^2.
\end{equation}

The Gamma function $\Gamma(-\ell)$ has simple poles at $\ell=m\in\mathbb{Z}_{\ge0}$, corresponding to the physical eigenvalues $\xi=E_m$. Near $\ell=m$,
\begin{equation}
\Gamma(-\ell) \sim \frac{(-1)^{m+1}}{m!(\ell-m)} .
\end{equation}
At the same time, the denominator $\Gamma(n+1-\ell)$ diverges for $n<m$,
implying that the resolvent amplitudes 
have subleading contributions for $n<m$, which become negligible for large $m$.
Therefore, at resonance the Krylov distribution has strict support only for
\begin{equation}
n\ge m.
\end{equation}

For $n\ge m$, all amplitudes diverge uniformly as $1/(\ell-m)$, which cancels
after normalization of the probability distribution $P_n$. The remaining
$n$-dependence is controlled by the factorial ratio and the confluent
hypergeometric function. When $\ell=m$ is an integer,
${}_1F_1(-m,n+1-m,\gamma^2)$ truncates to a Laguerre polynomial and grows only
polynomially with $n$. Consequently, the dominant large-$n$ behavior of the
normalized distribution is
\begin{equation}
P_n(E_m)\propto \frac{\gamma^{2n}}{n!},
\qquad n\ge m,
\end{equation}
up to subleading polynomial corrections that do not affect expectation values.

Introducing the shifted variable $k=n-m\ge0$, the distribution reduces to a
Poisson distribution,
\begin{equation}
P_{m+k}(E_m)
=
e^{-\gamma^2}\frac{\gamma^{2k}}{k!},
\end{equation}
which is normalized and has mean $\langle k\rangle=\gamma^2$. The Krylov
complexity at the $m$-th resonance is therefore
\begin{equation}
\mathcal D(E_m)
=
\sum_{k=0}^\infty (m+k)P_{m+k}
=
m+\gamma^2.
\end{equation}

This result admits a simple physical interpretation. The integer $m$ reflects
the minimal Krylov depth required to reach the $m$-th energy eigenstate, enforced
by the analytic structure of the resolvent. The additional contribution
$\gamma^2$ arises from the coherent-state dressing induced by the displacement
and represents the intrinsic width of the resolvent state in Krylov space.
Thus, at resonance the resolvent behaves as an eigenstate at level $m$ dressed
by a coherent cloud of size $\gamma^2$.

The linear growth of $\mathcal D(E_m)$ with $m$, together with the additive
shift by $\gamma^2$, is a distinctive feature of quadratic Hamiltonians with
unbounded Lanczos coefficients $b_n\propto\sqrt{n}$. It provides a controlled
benchmark illustrating how Krylov distribution delocalizes at spectral
resonances while remaining $\mathcal O(1)$ away from them.

To illustrate this general behavior discussed in the main text, we numerically compute the Krylov distribution; the resulting behavior is shown in Fig.~\ref{D-Har}.


\section{$SU(1,1)$ Chain: Exact Krylov Analysis}
\label{sec:su11_model}

We consider the dynamics of a quantum state in Krylov space, generated by a Hamiltonian whose matrix elements in the Krylov basis $\{|n\rangle\}_{n=0}^\infty$ follow from the Lanczos algorithm. The particular model we study has Lanczos coefficients
\begin{equation}
a_n = 0, \qquad b_n = \alpha\sqrt{n(n+h-1)}, \quad n\ge 0,
\end{equation}
with $h>0$ and $\alpha>0$ a constant with dimensions of frequency. The Hamiltonian in this basis acts as
\begin{equation}
H|n\rangle = b_n|n-1\rangle + b_{n+1}|n+1\rangle,
\end{equation}
with the convention $b_0=0$. This form corresponds to a tight-binding chain with position-dependent hopping. The time-evolved state starting from $|0\rangle$ is
\begin{equation}
|\psi(t)\rangle = e^{-iHt}|0\rangle = \sum_{n=0}^\infty \phi_n(t)\,|n\rangle,
\end{equation}
with initial condition $\phi_n(0)=\delta_{n0}$. From the Schrödinger equation $i\partial_t|\psi(t)\rangle = H|\psi(t)\rangle$, we obtain the coupled differential equations
\begin{equation}
i\dot{\phi}_n(t) = b_{n+1}\phi_{n+1}(t) + b_n\phi_{n-1}(t)\,.
\end{equation}

The form of $b_n$ suggests an underlying $SU(1,1)$ algebraic structure \cite{Balasubramanian:2022tpr}. Indeed, with $k = h/2$, we define operators
\begin{align}
L_+|n\rangle &= \sqrt{(n+1)(n+2k)}|n+1\rangle, \nonumber\\
L_-|n\rangle &= \sqrt{n(n+2k-1)}|n-1\rangle,
\end{align}
which satisfy $[L_-,L_+] = 2L_0$ and $[L_0,L_\pm] = \pm L_\pm$, where $L_0|n\rangle = (n+k)|n\rangle$. Therefore,
$H = \alpha(L_+ + L_-)$, and the initial state $|0\rangle$ is the lowest weight state of the discrete series representation $D_k^+$.
Thus to solve for $\phi_n(t)$, one may use the disentangling formula for $SU(1,1)$, which up to a phase is
\begin{equation}
e^{-i\alpha t(L_+ + L_-)} = e^{-\tanh(\alpha t)L_+}
\left(\cosh(\alpha t)\right)^{-2L_0} e^{-\tanh(\alpha t)L_-}\,.
\end{equation}
Applying this to $|0\rangle$ and noting $L_-|0\rangle=0$, one obtains
\begin{equation}
|\psi(t)\rangle = \frac{1}{\cosh^{2k}(\alpha t)}\sum_{n=0}^\infty \frac{(-\tanh(\alpha t)L_+)^n}{n!}|0\rangle.
\end{equation}
Using $L_+^n|0\rangle = \sqrt{\frac{(2k)_n}{n!}}|n\rangle$ and absorbing the sign into the phase, one finds
\cite{Parker:2018yvk, Balasubramanian:2022tpr}
\begin{equation}
\phi_n(t) = \sqrt{\frac{(h)_n}{n!}}\; \frac{\tanh^n(\alpha t)}{\cosh^{h}(\alpha t)}, \qquad n\ge 0,
\end{equation}
where $(h)_n = h(h+1)\cdots(h+n-1)$ is the Pochhammer symbol. These satisfy $\sum_{n=0}^\infty |\phi_n(t)|^2 = 1$ for all $t$.

To compute Krylov resolvent amplitudes we use the same Laplace transform convention as for the quadratic model:
\begin{equation}
\psi_n(\xi) = -i\sqrt{\frac{(h)_n}{n!}} \int_0^\infty dt\, e^{i\xi t} \frac{\tanh^n(\alpha t)}{\cosh^h(\alpha t)}.
\end{equation}
Make the change of variables $u = \tanh(\alpha t)$, with
\begin{equation}
dt = \frac{1}{\alpha}\frac{du}{1-u^2}, \quad \frac{1}{\cosh^{h}(\alpha t)} =
(1-u^2)^{\frac{h}{2}},
\end{equation}
yielding
\begin{equation}\label{psin-int}
\psi_n(\xi) = -\frac{i}{\alpha}\sqrt{\frac{(h)_n}{n!}} 
\int_0^1 u^n (1-u^2)^{\frac{h}{2}-1} \left(\frac{1+u}{1-u}\right)^{\frac{i\xi}{2\alpha}} du.
\end{equation}
Define the dimensionless spectral parameters
\begin{equation}
a =  \frac{i\xi}{2\alpha} + \frac{h}{2}, \quad b = - \frac{i\xi}{2\alpha} + \frac{h}{2}.
\end{equation}
Then the integral can be rewritten as
\begin{equation}
\psi_n(\xi) = -\frac{i}{\alpha}\sqrt{\frac{(h)_n}{n!}} 
\int_0^1 u^n (1-u)^{a-1} (1+u)^{b-1} du.
\label{eq:psi_integral_su11}
\end{equation}
This expression can be expressed in terms of the Gauss hypergeometric function:
\begin{align}
\psi_n(\xi) =& -\frac{i}{\alpha}
\frac{\Gamma(a) \sqrt{(h)_n n!}}{\Gamma(n+a+1)} \nonumber\\
&\times {}_2F_1\!\left(1-b,\; n+1;\; n+a+1;\; -1\right).
\label{eq:psi_su11_exact}
\end{align}
Using known properties of the hypergeometric function ${}_2F_1$, one can verify that these Krylov resolvent amplitudes satisfy the expected recursion relation
\begin{equation}
i\xi\psi_n(\xi) = b_{n+1}\psi_{n+1}(\xi) + b_n\psi_{n-1}(\xi), \quad n\ge 1. \label{eq:psin}
\end{equation}

The analytic structure of \eqref{eq:psi_su11_exact} differs fundamentally from the quadratic model studied in the previous appendix. The Gamma function $\Gamma(a)$ has poles when $a=-m$ for $m=0,1,2,\dots$, giving
\begin{equation}
\xi_m = i\alpha(2m + h).
\end{equation}
These poles occur at imaginary $\xi$, not on the real axis. They correspond to decaying modes rather than stable energy eigenstates. This contrasts with the quadratic case where $\Gamma(-\ell)$ had poles at $\ell = m \in \mathbb{Z}_{\ge 0}$, giving real $\xi_m = \omega m - \gamma^2\omega$ (the actual energy eigenvalues). Indeed, the $SU(1,1)$ Hamiltonian $H = \alpha(L_+ + L_-)$ has continuous spectrum $(-\infty, \infty)$ when acting on the appropriate Hilbert space (the full discrete series representation). The initial state $|0\rangle$ is not an energy eigenstate but rather a wavepacket that spreads indefinitely.

To study the asymptotic behavior of the Krylov resolvent amplitudes for large $n$, it is useful to work directly with the integral representation \eqref{psin-int}. Due to the factor $u^n$, the integrand decays exponentially away from $u=1$, so the main contribution comes from the region near the upper bound. To systematically analyze this region, define $v = 1-u$ with $v \in [0,\delta]$ for $\delta \ll 1$. In terms of this variable and assuming real $\xi$ (so that $\kappa = \xi/(2\alpha) \in \mathbb{R}$), we have
\begin{align}
u^n &\sim e^{-nv}, \nonumber\\
(1-u^2)^{\frac{h}{2}-1} &\sim (2v)^{\frac{h}{2}-1},\nonumber \\
\left(\frac{1+u}{1-u}\right)^{i\kappa} &\sim \left(\frac{2}{v}\right)^{i\kappa}.
\end{align}
Thus the integral in \eqref{psin-int} becomes
\begin{equation}
\psi_n(\xi) \approx -\frac{i}{\alpha}\sqrt{\frac{(h)_n}{n!}} 2^{\frac{h}{2}-1 + i\kappa}
\int_0^{\delta} e^{-nv} v^{\frac{h}{2}-1 - i\kappa} \left[1 + O(v)\right] dv.
\end{equation}
Setting $t = nv$, we obtain
\begin{equation}
\psi_n(\xi) \approx -\frac{i}{\alpha}\sqrt{\frac{(h)_n}{n!}} 2^{\frac{h}{2}-1 + i\kappa}
 n^{-\frac{h}{2} + i\kappa} 
\int_0^{n\delta} e^{-t} t^{\frac{h}{2}-1 - i\kappa} dt.
\end{equation}
For $n \to \infty$, the upper limit $n\delta \to \infty$ (taking $\delta$ fixed or slowly decreasing with $n$), so
\begin{equation}
\int_0^{n\delta} e^{-t} t^{\frac{h}{2}-1 - i\kappa} dt = 
\Gamma\left(\frac{h}{2} - i\kappa\right) + O\left(e^{-n\delta/2}\right).
\end{equation}
Thus
\begin{equation}
\psi_n(\xi) \approx -\frac{i}{\alpha}\sqrt{\frac{(h)_n}{n!}} 2^{\frac{h}{2}-1 + i\kappa}
\Gamma\left(\frac{h}{2} - i\kappa\right) n^{-\frac{h}{2} + i\kappa}.
\end{equation}

Now we need the large-$n$ behavior of the prefactor $\sqrt{(h)_n/n!}$. Using Stirling's formula,
\begin{align}
(h)_n &= \frac{\Gamma(n+h)}{\Gamma(h)} \\
&\sim \frac{n^{h-1} \Gamma(n+1)}{\Gamma(h)} \left[1 + \frac{h(h-1)}{2n} + O(n^{-2})\right],
\end{align}
we obtain
\begin{equation}
\sqrt{(h)_n/n!} \sim \frac{n^{(h-1)/2}}{\sqrt{\Gamma(h)}} \left[1 + \frac{h(h-1)}{4n} + O(n^{-2})\right].
\end{equation}
Substituting this into the expression for $\psi_n(\xi)$ yields
\begin{equation}
\psi_n(\xi) \approx D(\xi,h) \, n^{-1/2 + i\kappa} \left[1 + O(n^{-1})\right],
\end{equation}
where
\begin{equation}
D(\xi,h) = -\frac{i}{\alpha\sqrt{\Gamma(h)}} 2^{\frac{h}{2}-1 + i\kappa}
\Gamma\left(\frac{h}{2} - i\kappa\right).
\end{equation}
This asymptotic form allows us to determine the large-$n$ behavior of the Krylov distribution. We find
\begin{equation}
|\psi_n(\xi)|^2 \sim \frac{K(\xi,h)}{n} \quad \text{as } n \to \infty, \quad \xi \in \mathbb{R},
\label{eq:main_result}
\end{equation}
where $K(\xi,h) = |D(\xi,h)|^2$. This power-law decay has immediate consequences for the normalization sum and Krylov complexity. In particular,
\begin{equation}
\sum_{n=0}^N |\psi_n(\xi)|^2 \sim K(\xi,h) \ln N \quad \text{for large }N,
\end{equation}
and the Krylov distribution exhibits the asymptotic scaling
\begin{equation}
\mathcal{D}_N(\xi) \equiv \frac{\sum_{n=0}^N n\,|\psi_n(\xi)|^2}{\sum_{n=0}^N |\psi_n(\xi)|^2} \sim \frac{N}{\ln N} \quad \text{as } N \to \infty.
\end{equation}

The $SU(1,1)$ chain thus provides an example in which the Krylov distribution grows indefinitely with the cutoff $N$, without exhibiting discrete resonances. This behavior stands in sharp contrast to the quadratic model, where at discrete energies $E_m$ one finds sharply defined peaks with $\mathcal{D}(E_m)=m+\gamma^2$. The absence of such resonances in the $SU(1,1)$ model can be traced to several fundamental differences. First the operator $H=\alpha(L_+ + L_-)$ acts as a hyperbolic generator of $SU(1,1)$, leading to unbounded motion and a purely continuous spectrum, rather than the effectively confining dynamics of the quadratic Hamiltonian. Second, the initial state $|0\rangle$ is the lowest-weight state of the $SU(1,1)$ discrete series but is not an energy eigenstate; under time evolution it forms a wavepacket that spreads indefinitely rather than undergoing periodic motion. These features are reflected mathematically in the power-law tail $|\psi_n(\xi)|^2 \sim n^{-1}$ for real $\xi$, which is a characteristic signature of continuous spectrum in Krylov space. By contrast, systems with slower Lanczos growth, such as $b_n\sim\sqrt{n}$ in the quadratic model, typically exhibit discrete spectra and exponentially localized resolvent amplitudes.

It is important to emphasize that this behavior also differs from the generic expectations for orthogonal polynomials discussed in Section~III. The departure from standard asymptotics arises because the Lanczos coefficients in the $SU(1,1)$ chain are unbounded and grow linearly, $b_n\sim \alpha n$, at large $n$. Such linear growth places the model outside the usual Nevai class and qualitatively modifies the asymptotic properties of the associated orthogonal polynomials, resulting in logarithmically divergent normalization and the continuous-spectrum Krylov scaling observed here.

\bibliography{literature}

\end{document}